\documentclass[twocolumn]{autart}
\usepackage[T1]{fontenc}
\usepackage[latin9]{inputenc}
\usepackage{color}
\usepackage{textcomp}
\usepackage{mathrsfs}
\usepackage{amsmath,epsfig,epstopdf}
\usepackage{amssymb}
\usepackage{graphicx}
\usepackage{esint}

\def\Real{\ensuremath{\mathbb{R}}}
\newtheorem{asm}[thm]{Assumption}
\newtheorem{prb}[thm]{Problem}

\begin{document}

\begin{frontmatter}

\title{Nonlinear Control of a Tethered UAV:\\ the Taut Cable case\thanksref{footnoteinfo}}

\thanks[footnoteinfo]{This work is supported by a FRIA scholarship grant and the PF7 European project SHERPA.}

\author[ULB,Unibo]{Marco M. Nicotra}\ead{mnicotra@ulb.ac.be},
\author[Unibo]{Roberto Naldi}\ead{roberto.naldi@unibo.it},
\author[ULB]{Emanuele Garone}\ead{egarone@ulb.ac.be}

\address[ULB]{Service d'Automatique et d'Analyse des Syst\`emes, Universit\'e Libre de Bruxelles\\Av. F.D. Roosevelt 50, CP 165/55, 1050 - Bruxelles, Belgium}  
\address[Unibo]{Center for Research on Complex Automated Systems, Alma Mater Studiorum (University of Bologna)\\
Viale C. Pepoli 3/2 40136 - Bologna, Italy}             


\begin{abstract}
This paper focuses on the design of a stabilizing control law for
an aerial vehicle which is physically connected to a ground station
by means of a tether cable. By taking advantage of the tensile force
acting along the taut cable, it is shown that the tethered UAV is able
to  maintain a non-zero attitude while hovering in a constant position.
The control objective is to stabilize the desired configuration while
simultaneously ensuring that the cable remains taut at all times. This
leads to a nonlinear control problem subject to constraints. This paper
provides a two-step solution. First, the system is stabilized using a
cascade control scheme based on thrust vectoring. Then, constraint
satisfaction is guaranteed using a novel Reference Governor scheme.
\end{abstract}

\end{frontmatter}

\section{INTRODUCTION}

Recent advancements in the field of Unmanned Aerial Vehicles (UAVs)
have lead to the availability of inexpensive aerial robots with a
growing range of applications ranging from surveillance \cite{UAV_CoopSurveillance}
to advanced robotic operations including environment interaction \cite{UAV_ForcePositionFeedback},
grasping \cite{UAV_Grasper}-\cite{UAV_CoopGrasping} and manipulation \cite{UAV_CoopConstruction}.
The full potential of these systems, however, is still limited
by key factors such as flight time, computing capabilities and airspace safety
regulations \cite{UAV_BookChapter}. A possible solution to these limitations is to connect the UAV to a ground station
by means of a tether cable able to supply energy, transmit data and/or apply forces.

Since the dynamic properties of the UAV are deeply influenced by the cable,
the safe deployment of tethered UAVs requires the development of specific
control strategies.
Early works on the subject \cite{UAVTether_Eigenfunction}-\cite{UAVTether_LongitudinalMotion}
studied the stabilization of tethered UAVs using linearized models.
Although the primary interest in tethered UAVs is their virtually unlimited
flight-time \cite{UAVTether_OilDetection}, recent results have shown the
advantage of using the taut cable as an additional control input.
Possible examples include: guiding the landing
of a helicopter on a ship \cite{UAVTether_Helilanding}, improving
fight stability in the presence of wind \cite{UAVTether_HoveringStability},\cite{M_UAVKiteConfig},
and using multiple cables to achieve full actuation \cite{UAVGroundCoop_ControlAllocation}.
Moreover, it has been shown in \cite{UAVTether_Sensing},\cite{UAVTether_Geometric}
that the taut cable configuration can also be used to measure the position of the UAV. A
common feature of these papers is that the cable tension is controlled
by an actuated winch, whereas the UAV position is controlled by the
UAV itself.

This paper investigates an alternative approach where the actuated winch imposes
only the cable length whereas the UAV controls its elevation angle while
ensuring a minimal cable tension. It is worth noting that the proposed
control law can also be applied to the case of a fixed-length cable
since it does not require the presence of an actuated winch. To the author's best knowledge, this approach
to the control of a tethered UAV has not been addressed previously.

The first contribution of the paper is to show that the tethered UAV is able to
achieve a set of equilibrium configurations that is different from the
untethered case. This set is characterized both analytically and geometrically.
The main contribution of this paper is the development of an ad-hoc strategy for ensuring
constraint satisfaction at all times. The proposed solution consists in two
separate design steps: first, the nonlinear system is stabilized using a cascade
approach \cite{UAV_Review}. Second, the closed loop system is augmented with a specifically designed
Reference Governor (RG) that ensures constraint enforcement by introducing a series
of intermediate waypoints. Although several RG strategies exist in the literature
(see \cite{RefGov_Survey} and references therein), the proposed methods are not well suited
for the present application. As such, the paper proposes a novel backtracking RG
strategy that generates the waypoint sequence off-line to avoid computationally
intensive on-line operations. To do so, particular effort has been dedicated to the
characterization of the set invariance of the closed loop system.

A preliminary conference version of this paper appeared in \cite{M_UAVTautCable}.
The main novelty with respect to this earlier work is the introduction of the
backtracking RG algorithm. Other major improvements include the analytical
characterization of the set of attainable steady-state attitudes, more rigourous stability proofs and the
determination of a more stringent inner loop gain using the $\ell_1$ norm.

\section{PRELIMINARIES}
This section provides a brief description of the notation that will be used throughout the paper. In particular, let $\mathbb{R}_{>0}$ denote the set $\{x\in\mathbb{R}:\;x>0\}$, let $\mathbb{R}_{\geq0}$ denote the set $\{x\in\mathbb{R}:\;x\geq0\}$, let $\left\Vert\cdot\right\Vert$ denote the Euclidean norm, and let $\left\Vert\cdot\right\Vert_\infty$ denote the infinity norm as in \cite{Isidori}.
Moreover, define the saturation function $\sigma_{\lambda}\left(x\right)$
\[
\sigma_{\lambda}\left(x\right)=\mathrm{sign}(x)\min\left(\left\vert x\right\vert,\lambda\right)
\]
and the $\mathrm{atan2}\left(y,x\right)$ function
\[
\mathrm{atan2}\left(y,x\right)=\left\{\begin{array}{ll}
\arctan \frac{y}{x} & x>0 \\
\arctan \frac{y}{x}+\pi & y\geq0,\,x>0 \\
\arctan \frac{y}{x}-\pi & y<0,\,x>0 \\
\:\frac{\pi}{2} & y>0,\,x=0 \\
-\frac{\pi}{2} & y<0,\,x=0 \\
\mathrm{undefined} & y=0,\,x=0
\end{array}\right.
\]
The following definition of Input-to-State Stability (ISS) given in \cite{ISS_BasicConcepts} is reported for the sake of completeness.

\begin{defn}
A system $\dot{x}=f\left(x,u\right)$ with $x\in\mathbb{R}^n$ and $u\in\mathbb{R}$ is Input-to-State Stable (ISS) with restriction $\mathcal {X}\subset\mathbb{R}^n$ on the initial state $x(0)$ and restriction $\mathcal{U}\subset\mathbb{R}$ on the input $u$ if there exist a class-$\mathcal{K}$ function\footnote{A continuous function $\gamma(x)$ is said to be of class-$\mathcal{K}$ if it is strictly increasing and satisfies $\gamma(0)=0$.} $\gamma:\mathbb{R}\rightarrow\mathbb{R}$ and a class-$\mathcal{KL}$ function\footnote{A continuous function $\beta(x,s)$ is said to be of class-$\mathcal{KL}$ if, for each fixed $s$, $\beta(x,s)$ is a class-$\mathcal{K}$ function and, for each fixed $x$, $\beta(x,s)$ is decreasing and satisfies $\beta(x,s)\rightarrow0$ for $s\rightarrow\infty$.} $\beta:\mathbb{R}^2\rightarrow\mathbb{R}$ such that
\begin{equation}\label{eq: ISS}
\left\Vert x\left(t\right) \right\Vert \leq \beta\left(\left\Vert x\left(0\right) \right\Vert,t\right)+\gamma\left(\sup_{\tau\leq t}\left\Vert u\left(\tau\right) \right\Vert\right),
\end{equation}
for all $x\left(0\right)\in\mathcal{X}$ and $u\left(t\right)\in\mathcal{U}$.
\end{defn}

\section{PROBLEM STATEMENT}\label{sec: PROBLEM STATEMENT}

\subsection{System Modeling}
\begin{figure}
\includegraphics[scale=0.72]{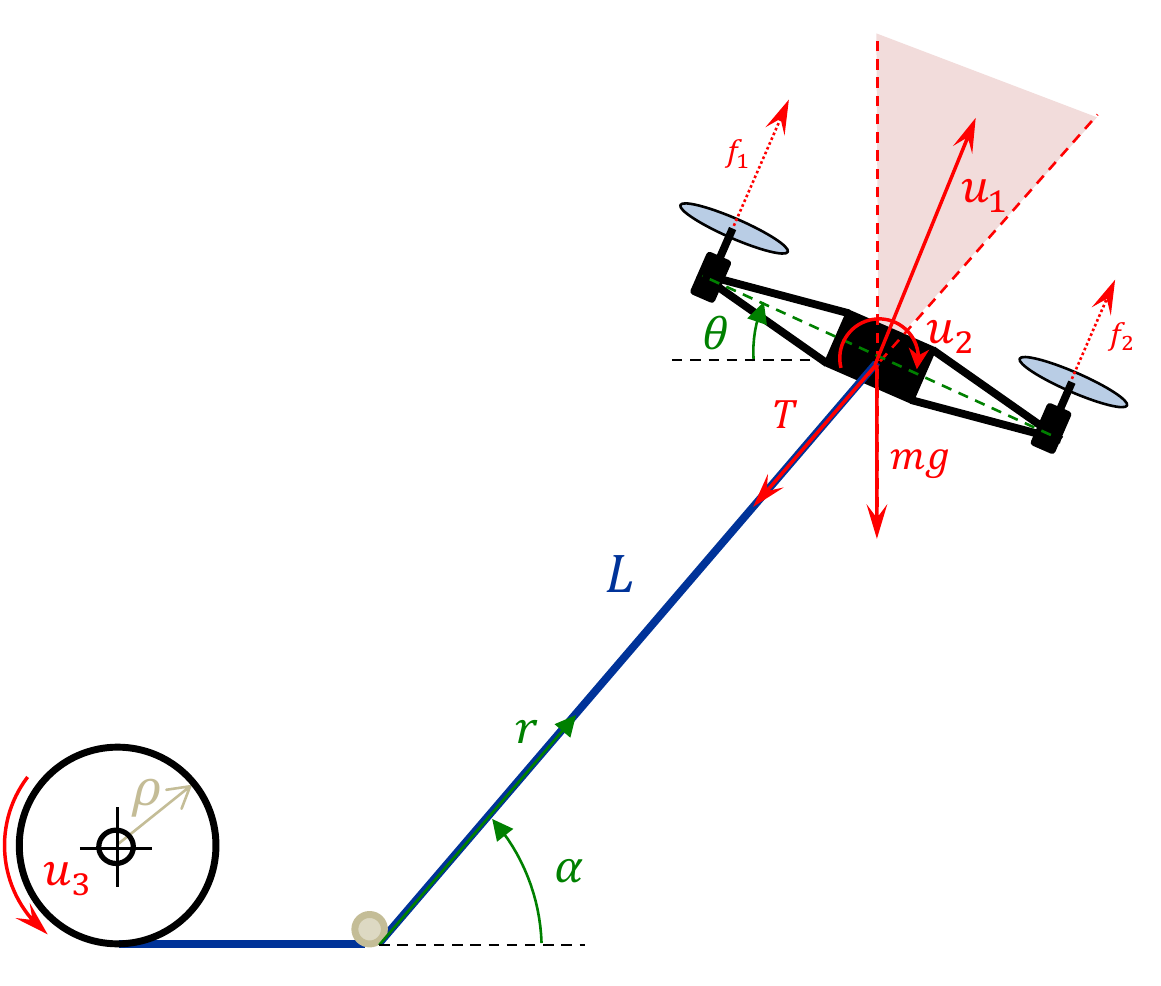}
\caption{\label{fig: UAV model}Planar model of a tethered UAV with
a taut cable}
\end{figure}

Consider the planar model of a tethered UAV depicted in Figure \ref{fig: UAV model}. The vehicle has mass $m\in\mathbb{R}_{>0}$, moment of inertia $\mathcal{J}\in\mathbb{R}_{>0}$ and is physically connected to the ground by means of a tether cable of length $L\in\mathbb{R}_{>0}$.
Let the \textit{radial position} $r\in\mathbb{R}_{>0}$ and the \textit{elevation angle} $\alpha\in\left[0,\pi\right]$ be the polar coordinates of the UAV, and let the \textit{pitch angle} $\theta\in\left(-\pi,\pi\right]$ be the attitude of the UAV with respect to the horizon.\\
The vehicle is subject to the gravity acceleration $g$, the cable tension $T \in \Real_{\geq 0}$, and is actuated by two propellers that generate a total thrust $u_1 \in \Real_{\geq 0}$ and a resultant torque $u_2 \in \Real$. The UAV actuator dynamics are assumed to be negligible. The cable is governed by a control torque $u_3\in \Real$ that acts on a winch of radius $\rho\in \Real_{\geq 0}$ and moment of inertia ${\mathcal I}\in \Real_{\geq 0}$.
The following approximations are made.

\begin{asm}\label{asm: cable}
The cable is inextensible, massless and has zero shear stiffness. Moreover, it is attached to the center of mass of the UAV.
\end{asm}
\begin{asm}\label{asm: air}
Air viscosity is negligible.
\end{asm}
Under Assumption \ref{asm: cable}, the total kinetic energy $\mathcal{K}$ and potential energy $\mathcal{P}$ of the UAV are
\[
\begin{array}{l}
\mathcal{K}=\frac{1}{2}\frac{\mathcal{I}}{\rho^2}\dot{L}^2+\frac{1}{2}m\dot{r}^{2}+\frac{1}{2}mr^{2}\dot{\alpha}^{2}+\frac{1}{2}\mathcal{J}\dot{\theta}^{2}\\
\mathcal{P}=mgr\sin\alpha.
\end{array}
\]
Following from Assumption \ref{asm: air}, it is possible to define the Lagrangian function $\mathscr{\mathcal{L}=\mathcal{K}-\mathcal{P}}$.
The dynamic model of the system can then be obtained via the Euler-Lagrange theorem
\[
\frac{d}{dt}\frac{\partial\mathcal{L}}{\partial\dot{q}_{i}}-\frac{\partial\mathcal{L}}{\partial q_{i}}=F_{i}\qquad i=L,r,\alpha,\theta
\]
where
\[
\begin{array}{ll}
F_L=u_3+\rho T, &
F_{r}=u_{1}\sin\left(\alpha+\theta\right)-T,\\
F_{\alpha}=ru_{1}\cos\left(\alpha+\theta\right), &
F_{\theta}=u_{2}.
\end{array}
\]
This leads to the dynamic model
\begin{equation}\label{eq: General System Dynamics}
\left\{ \begin{array}{l}
\frac{\mathcal{I}}{\rho}\ddot{L}=u_3+\rho T\\
m\ddot{r}=mr\dot{\alpha}^{2}-mg\sin\alpha+u_{1}\sin\left(\alpha+\theta\right)-T\\
mr^{2}\ddot{\alpha}=-2mr\dot{r}\dot{\alpha}-mgr\cos\alpha+ru_{1}\cos\left(\alpha+\theta\right)\\
\mathcal{J}\ddot{\theta}=u_{2},
\end{array}\right.
\end{equation}
which is the generic model for a tethered UAV. To specialize it to the taut cable configuration, the following definition is given
\begin{defn}\label{def: TautCable}
The cable is taut at time $t$ if $r(t)=L(t)$.
\end{defn}
Due to the unilateral nature of the cable forces (i.e. its inability to withstand compression), it follows from Assumption \ref{asm: cable} that the cable remains taut if its tension remains always positive, i.e. $T>0$, where the cable tension $T$ is
\begin{equation}\label{eq: Tension}
T\left(r,\alpha,\theta\right)=mr\dot{\alpha}^{2}-mg\sin\alpha+u_{1}\sin\left(\alpha+\theta\right)-m\ddot{r}.
\end{equation}
Assuming that the cable is taut $\forall t\geq0$, the dynamic model (\ref{eq: General System Dynamics}) of the tethered UAV can be rewritten as
\begin{equation}\label{eq: Taut Cable Dynamics}
\left\{ \begin{array}{l}
\ddot{r}=\frac{\rho}{\mathcal{I}}u_{3}+\frac{\rho^2}{\mathcal{I}}T\\
\ddot{\alpha}=-\frac{1}{r}\left(2\dot{r}\dot{\alpha}+g\cos\alpha\right)+\frac{1}{mr}u_{1}\cos\left(\alpha+\theta\right)\\
\ddot{\theta}=\frac{1}{\mathcal{J}}u_{2}
\end{array}\right.
\end{equation}
subject to the constraint
\begin{equation}\label{eq: Taut Cable Constraint}
T\left(r,\alpha,\theta\right)>0.
\end{equation}
\begin{rem}\label{rmk: CableAssumption}
It is worth noting that if Assumption \ref{asm: cable} is dropped, the general approach presented in this paper remains valid with minor modifications. Notably, the cable weight and inertia must be added to the $\ddot{\alpha}$ dynamics in equation \eqref{eq: Taut Cable Dynamics}. Moreover, in the presence of a non-zero mass and a non-infinite stiffness, the taut cable definition must be changed to $r(t)\geq L(t)$. Given this new definition, it is possible to compute (or determine experimentally) a minimum cable tension $T_{\min}$ such that $T>T_{\min}$ ensures the taut cable condition. Please note that for cables with high stiffness and low mass, it is typically reasonable to assume $L\approx r$ whenever $T>T_{\min}$.
\end{rem}

\subsection{Control Objectives}
The objective of this paper is to stabilize the tethered UAV dynamics  (\ref{eq: Taut Cable Dynamics}) to a constant reference while simultaneously satisfying the taut cable constraint (\ref{eq: Taut Cable Constraint}). To do so, it is required that the desired reference must be \textit{attainable} as per the following definition.

\begin{defn}\label{def: Attainable Set}
\textbf{Attainable Equilibria:} Given a safety margin $\epsilon\geq0$, the set of steady-state admissible configurations $\mathcal{S}_{\epsilon}$
is the set of equilibrium points $\left(\bar{r},\bar{\alpha},\bar{\theta}\right)$ such that $\bar{T}:=T\left(\bar{r},\bar{\alpha},\bar{\theta}\right)>\epsilon$.
\end{defn}
Using this definition, the control objectives can be stated as follows.

\begin{prb}\label{prb: ControlObj}
\textit{Given a set-point $\left(\bar{r},\bar{\alpha},\bar{\theta}\right)\in\mathcal{S}_{\epsilon}$,
design a control law such that}
\begin{equation} \label{eq: Obj_Stability}
\underset{t\rightarrow\infty}{\lim}\left(r\left(t\right),\alpha\left(t\right),\theta\left(t\right)\right)=\left(\bar{r},\bar{\alpha},\bar{\theta}\right)
\end{equation}
\begin{equation}\label{eq: Obj_Constraint}
T\left(r\left(t\right),\alpha\left(t\right),\theta\left(t\right)\right)>0,\quad\forall t\geq0.
\end{equation}
\end{prb}

\section{ATTAINABLE SETPOINTS}\label{sec: ATTAINABLE EQUILIBRIUM POINTS}

\begin{figure*}
\includegraphics[scale=0.88]{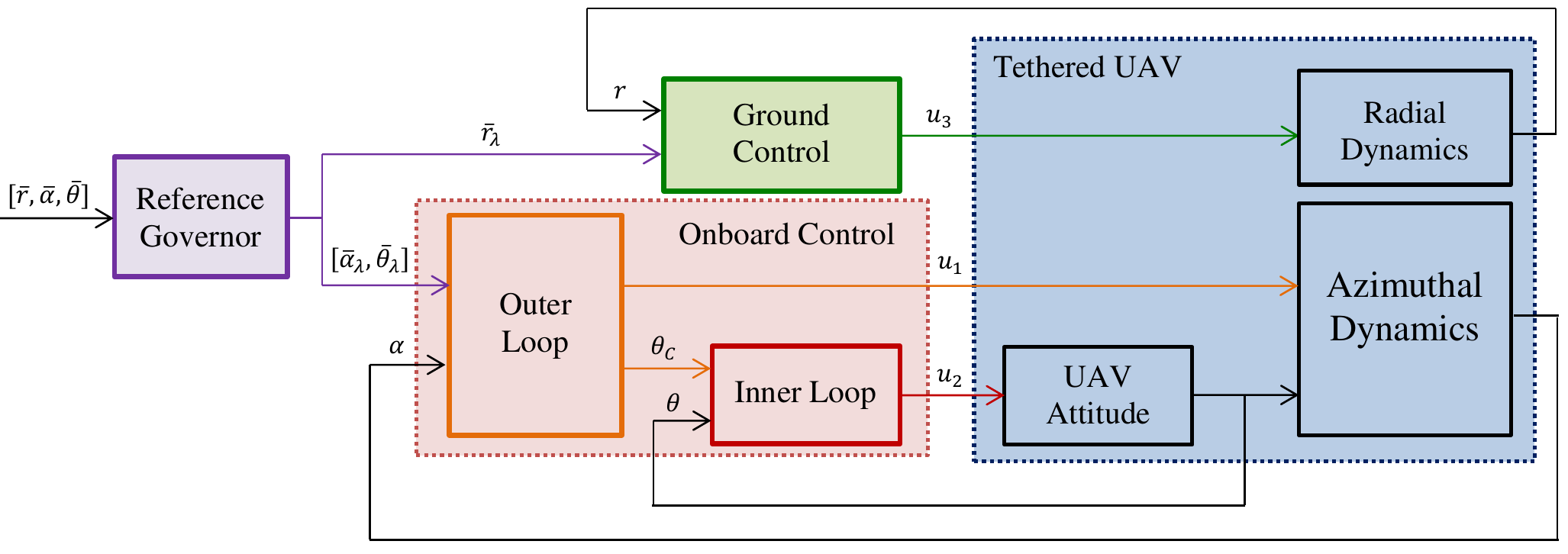}\caption{\label{fig: Control architecture}Proposed control architecture.}
\end{figure*}
The goal of this section is to compute the set $\mathcal{S}_{\epsilon}$.
\begin{prop}\label{prop: Attainable Set}
Let system (\ref{eq: Taut Cable Dynamics}) be subject to constraint (\ref{eq: Taut Cable Constraint}). The set of attainable equilibria $\mathcal{S}_{\epsilon}$ consists of all $\left(\bar{r},\bar{\alpha},\bar{\theta}\right)$ satisfying
\begin{equation}
\begin{array}{l}
\begin{array}{c}
\begin{array}{ccc}
\bar{r}>0, & \quad & \bar{\alpha}\in[0,\pi]\end{array},\\
\quad\\
\left\{ \begin{array}{ll}
\bar{\theta}\in\left(\bar{\theta}_\epsilon\left(\bar\alpha\right),\:\frac{\pi}{2}-\bar\alpha\right) & \mathrm{if}\:\bar{\alpha}\in\left[0,\frac{\pi}{2}\right)\\
\bar{\theta}=0 & \mathrm{if}\:\bar{\alpha}=\frac{\pi}{2}\\
\bar{\theta}\in\left(\frac{\pi}{2}-\bar\alpha,\:\bar{\theta}_\epsilon\left(\bar\alpha\right)\right) & \mathrm{if}\:\bar{\alpha}\in\left(\frac{\pi}{2},\pi\right]
\end{array}\right.
\end{array}\end{array}\label{eq: Attainable Attitudes}
\end{equation}
with
\begin{equation}\label{eq: BetaLim}
\bar{\theta}_\epsilon\left(\bar\alpha\right)=\mathrm{atan}\left(\frac{\epsilon}{mg\cos\bar{\alpha}}+\tan\bar{\alpha}\right)-\bar\alpha.
\end{equation}
Moreover, the cable tension at equilibrium is
\[
\bar{T}=\left\{ \begin{array}{ll}
\mathrm{any}\,\mathbb{R}_{>0} & \mathrm{if}\:\bar{\alpha}=\frac{\pi}{2}\\
mg\left(\tan\left(\bar\alpha+\bar\theta\right)\cos\bar{\alpha}-\sin\bar{\alpha}\right) & \mathrm{if}\:\bar{\alpha}\in\left[0,\pi\right]\setminus\left\{ \frac{\pi}{2}\right\} .
\end{array}\right.
\]
\end{prop}
\begin{pf}
System (\ref{eq: Taut Cable Dynamics}) is at equilibrium for $u_2=\bar{u}_{2}:=0$, $u_3=\bar{u}_{3}:=-\rho \bar{T}$ and $u_1=\bar{u}_{1}$, where
\begin{equation}\label{eq: Equilibrium Input1}
\bar{u}_{1}\cos\left(\bar\alpha+\bar\theta\right)=mg\cos\bar{\alpha}.
\end{equation}
Following from (\ref{eq: Tension}) evaluated at steady state, condition $\bar{T}>\epsilon$ leads to
\begin{equation}\label{eq: SteadyState Constraint}
\bar{T}=\bar{u}_{1}\sin\left(\bar\alpha+\bar\theta\right)-mg\sin\bar{\alpha}>\epsilon.
\end{equation}
Depending on the value of $\bar{\alpha}$, two cases must be considered.\\
\textit{Case 1.} If $\bar{\alpha}=\frac{\pi}{2}$, equations (\ref{eq: Equilibrium Input1}) and (\ref{eq: SteadyState Constraint}) become
\[
\left\{ \begin{array}{l}
\bar{u}_{1}\cos\left(\bar\alpha+\bar\theta\right)=0\\
\bar{u}_{1}\sin\left(\bar\alpha+\bar\theta\right)> mg+\epsilon
\end{array}\right.
\]
which is verified for $\theta=0$ and $\forall\bar{u}_{1}> mg+\epsilon$.\\
\textit{Case 2.} If $\bar{\alpha}\neq\frac{\pi}{2}$, condition (\ref{eq: Equilibrium Input1})
is satisfied for
\begin{equation}
\bar{u}_{1}=mg\frac{\cos\bar{\alpha}}{\cos\left(\bar\alpha+\bar\theta\right)}\label{eq: u1Eq}
\end{equation}
which exists if and only if $\bar\alpha+\bar\theta\neq\pm\frac{\pi}{2}$.
Substituting (\ref{eq: u1Eq}) in (\ref{eq: SteadyState Constraint}), the taut cable condition becomes
\[
\bar{T}=mg\cos\bar{\alpha}\tan\left(\bar\alpha+\bar\theta\right)-mg\sin\bar{\alpha}>\epsilon
\]
which can be rewritten as
\begin{equation}\label{eq: SetEps}
\begin{array}{ll}
\tan\left(\bar\alpha+\bar\theta\right)>\frac{\epsilon}{mg\cos\bar{\alpha}}+\tan\bar{\alpha} & \mathrm{if}\:\bar{\alpha}\in\left[0,\frac{\pi}{2}\right)\\
\tan\left(\bar\alpha+\bar\theta\right)<\frac{\epsilon}{mg\cos\bar{\alpha}}+\tan\bar{\alpha} & \mathrm{if}\:\bar{\alpha}\in\left(\frac{\pi}{2},\pi\right].
\end{array}
\end{equation}
The solution of this inequality is given in equations (\ref{eq: Attainable Attitudes})-(\ref{eq: BetaLim}), which concludes the proof. $\hfill\Box$
\end{pf}

\begin{rem}
The largest set of attainable equilibria can be obtained by choosing $\epsilon=0$. Following from (\ref{eq: BetaLim}), this implies $\bar\theta_0(\bar{\alpha})=0$. As a result, for any given elevation angle $\bar{\alpha}$, the attitude $\bar{\theta}$ must ensure that the thrust vector is contained in the conic combination of the tension vector $T$ and the weight vector $mg$. This geometrical interpretation is depicted in Figure \ref{fig: UAV model}.
\end{rem}

\begin{rem}
Note that being able to maintain a non-zero attitude angle while hovering is a relevant feature for practical applications. Indeed, by changing the attitude of the UAV it is possible to direct onboard hardware (e.g. a camera) without the need of actuated joints. This can be beneficial in terms of both structural simplicity  and payload capacity.
\end{rem}

\section{CONTROL ARCHITECTURE}\label{sec:CONTROL STRATEGY}
\vspace{-2 mm}
The goal of this section is to describe the overall control strategy which will be developed in this paper. The proposed approach consists in pre-stabilizing the system dynamics and then using a reference governor to ensure constraint satisfaction by suitably manipulating the applied reference.\\

The pre-stabilizing control consists in a \textbf{ground control} unit, which imposes the radial position $r$, and an \textbf{onboard control} unit which is implemented directly on the UAV. In particular, the onboard control unit is based on a hierarchical cascade approach \cite{UAV_CascadeControl} where the \textbf{inner loop} controls the attitude dynamics $\theta$ and the \textbf{outer loop} controls the azimuth angle $\alpha$. Under the assumption that the inner loop is ideal, the outer loop is designed to ensure asymptotic stability while simultaneously enforcing the taut cable constraint.\\

After lifting the assumption on the inner loop, the stability of the inner/outer loop interconnection is proven with the aid of the small gain theorem. However, it will be shown that the transient dynamics may lead to a constraint violation if the desired reference is too far from the initial conditions. To solve this problem, the applied reference will issued by a \textbf{reference governor} which, if necessary, provides a succession of intermediate waypoint references so as to limit the transient dynamics of the closed-loop system. The proposed control architecture is illustrated in Figure \ref{fig: Control architecture}.
\vspace{-2 mm}

\section{GROUND CONTROL}\label{sec: GROUND CONTROL}
The objective of the ground station is to control the radial position of the UAV by acting on the winch. Since unwinding the cable too quickly may lead to a loss of tension, the control law must ensure that $r\left(t\right)$, as well as its first and second derivatives, remain bounded. To do so, a nested saturation control law is proposed
\begin{equation}
u_{3}=-\frac{\mathcal{I}}{\rho}\sigma_{\lambda_{1}}\left(k_{Dr}\dot{r}+\sigma_{\lambda_{2}}\left(k_{Pr}\left(r-\bar{r}\right)\right)\right)-\rho T.\label{eq: u3}
\end{equation}
The following proposition shows that $u_3$ as in (\ref{eq: u3}) is able to attain all of these objectives.
\begin{prop}\label{prop: Radial Control}
Given the radial dynamics
\begin{equation}
\ddot{r}=\frac{\rho}{\mathcal{I}}u_{3}+\frac{\rho^{2}}{\mathcal{I}}T\label{eq: Radial Dynamics}
\end{equation}
and control law (\ref{eq: u3}) with $k_{Dr}=2\sqrt{k_{Pr}}$, then for any $\bar{r} \geq 0$
\begin{enumerate}
\item the acceleration is bounded, i.e. $r\left(t\right)=\bar{r}$ is a Globally Asymptotically Stable (GAS) equilibrium point for any $\lambda_{1},\lambda_2>0$ and $k_{Pr}>0$;
\item $\left\Vert \ddot{r}\right\Vert _{\infty}\leq\lambda_{1}$;
\item if the initial velocity of $r$ satisfies $\left|\dot{r}\left(0\right)\right|\leq\frac{\lambda_{1}}{k_{Dr}}$, then:
\begin{itemize}
\item the velocity of $r$ will satisfy $\left\Vert \dot{r}\right\Vert _{\infty}\leq\frac{\max\left(\lambda_{1},\lambda_{2}\right)}{k_{Dr}}$
\item the trajectory of $r(t)$ is bounded by
\end{itemize}
\[
    r\left(t\right)\in\left[\min\left(\bar{r},r\left(0\right),r^\star\!\right),\max\left(\bar{r}, r\left(0\right),r^\star\!\right)\right],\;\forall t,
\]
where
\[
r^\star=\left\{\begin{array}{ll}
r(0)+\frac{\dot{r}(0)}{k_{Dr}}&\mathrm{if}\;\tilde{r}(0)\dot{r}(0)\geq0\\
\bar{r}+\Delta r^\star& \mathrm{if}\;\tilde{r}(0)\dot{r}(0)<0;\;\tau^\star>0\\
r(0) & \mathrm{otherwise}
\end{array}\right.
\]
and
\[
\tau^\star=\frac{\dot{r}(0)}{k_{Pr}\tilde{r}(0)+\sqrt{k_{Pr}}\dot{r}(0)}
\]
\[
\Delta r^\star=\left(r(0)-\bar{r}+\frac{\dot{r}(0)}{\sqrt{k_{Pr}}}\right)e^{-\sqrt{k_{Pr}}\tau^\star}.
\]
\end{enumerate}
\end{prop}

\begin{pf}
\begin{enumerate}
\item Define
\[
x_r=\left[\begin{array}{c} x_{r1} \\ x_{r2} \end{array}\right] = \left[\begin{array}{c} r-\bar{r} \\ \dot{r} \end{array}\right].
\]
The state-space equation of the controlled system is
\[
\left\{ \begin{array}{l}
\dot{x}_{r1}=x_{r2}\\
\dot{x}_{r2}=-\sigma_{\lambda_{1}}\left(k_{Dr}x_{r2}+\sigma_{\lambda_{2}}\left(k_{Pr}x_{r1}\right)\right)
\end{array}\right.
\]
which has the same form as the results in \cite{NestedSaturationControl_TVP-Robustness}. The origin of the controlled system is therefore GAS for any $\lambda_{1}, \lambda_2>0$, $k_{Pr}>0$ and $k_{Dr}>0$.
\item By definition of the saturation function, it follows that $\sigma_{\lambda1}(\cdot)\leq\lambda_{1}$.
\item See Appendix A. $\hfill\Box$
\end{enumerate}
\end{pf}

\section{ONBOARD CONTROL}\label{sec: UAV CONTROL}
The objective of the onboard control is to impose $\underset{t\rightarrow\infty}{\lim}\alpha\left(t\right)=\bar{\alpha}$
and $\underset{t\rightarrow\infty}{\lim}\theta\left(t\right)=\bar{\theta}$
without violating the taut cable constraint $T\left(r(t),\alpha(t),\theta(t)\right)>0$.

Since there are only two control inputs, $u_{1}$ and $u_{2}$, and
three control objectives, this problem could be ill-posed.
However, the following lemma shows that the control problem can be
achieved indirectly by satisfying two independent control objectives.
\begin{lem}
\label{lem: OuterLoop Objectives}
Consider system (\ref{eq: General System Dynamics})
and the constant reference $(\bar{r},\bar{\alpha},\bar{\theta})\in\mathcal{S}_{\epsilon}$. Given $\underset{t\rightarrow\infty}{\lim}r\left(t\right)=\bar{r}$,
the conditions
\[
\begin{array}{c}
\underset{t\rightarrow\infty}{\lim}\alpha\left(t\right)=\bar{\alpha}\\
\underset{t\rightarrow\infty}{\lim}\theta\left(t\right)=\bar{\theta}\\
T\left(r(t),\alpha(t),\theta(t)\right)>0
\end{array}
\]
are satisfied if
\begin{eqnarray}
\underset{t\rightarrow\infty}{\lim}\alpha\left(t\right)=\bar{\alpha} & \!\! & \!\!\label{eq: Control Objective1}\\
T=\bar{T}+mr\dot{\alpha}^{2} & \!\! & \!\!\label{eq: Control Objective2}
\end{eqnarray}
 with
\begin{equation}
\bar{T}=mg\left(\tan\left(\bar\alpha+\bar\theta\right)\cos\bar{\alpha}-\sin\bar{\alpha}\right).\label{eq: Equilibrium Tension}
\end{equation}
\end{lem}
\begin{pf}
Since reference $\left(\bar{r},\bar{\alpha},\bar{\theta}\right)\in\mathcal{S}_{0}$,
it follows from Proposition \ref{prop: Attainable Set} that $\bar{T}>0$.
As a result, $T=\bar{T}+mr\dot{\alpha}^{2}$ is strictly positive.
Furthermore, substituting the expression of $T$ and the property $\underset{t\rightarrow\infty}{\lim}\left[\alpha\left(t\right),r\left(t\right)\right]=\left[\bar{\alpha},\bar{r}\right]$,
it follows that
\[
\underset{t\rightarrow\infty}{\lim}\left\{ \begin{array}{l}
m\ddot{r}=-mg\sin\alpha+u_{1}\sin\left(\alpha+\theta\right)-\bar{T}\\
mr^{2}\ddot{\alpha}=-2mr\dot{r}\dot{\alpha}-mgr\cos\alpha+ru_{1}\cos\left(\alpha+\theta\right)
\end{array}\right.
\]
\begin{equation}
=\left\{ \begin{array}{l}
0=-mg\sin\bar{\alpha}+u_{1}\sin\left(\bar\alpha+\theta\right)-\bar{T}\\
0=-mg\bar{r}\cos\bar{\alpha}+\bar{r}u_{1}\cos\left(\bar\alpha+\theta\right)
\end{array}\right..\label{eq: Limit System}
\end{equation}
From the second equation of (\ref{eq: Limit System}) it follows that
\begin{equation}
\underset{t\rightarrow\infty}{\lim}u_{1}\left(t\right)=mg\frac{\cos\bar{\alpha}}{\cos\left(\bar\alpha+\theta\right)}.\label{eq: Limit Input}
\end{equation}
Substituting (\ref{eq: Equilibrium Tension}) and (\ref{eq: Limit Input})
into the first equation of (\ref{eq: Limit System}), it follows that
\[
\underset{t\rightarrow\infty}{\lim}mg\cos\bar{\alpha}\left(\tan\left(\bar\alpha+\theta\right)-\tan\left(\bar\alpha+\bar\theta\right)\right)=0.
\]
This implies $\underset{t\rightarrow\infty}{\lim}\theta\left(t\right)=\bar{\theta}$.
$\hfill\Box$ \end{pf}
Lemma \ref{lem: OuterLoop Objectives} provides a starting point for
the design of the hierarchical control architecture that will be developed
for the UAV. The remainder of this section is structured as follows. First,
the outer loop will be designed to satisfy all of the control objectives
(\ref{eq: Control Objective1})-(\ref{eq: Control Objective2})
under the assumption that the UAV attitude can be imposed instantaneously.
Then, the inner loop will be charged with pursuing the desired attitude in
such way that the stability of the inner/outer loop interconnection is not
compromised.

\subsection{Outer Loop Control}\label{ssec: Outer Loop}
Assuming that the UAV attitude can be imposed instantaneously, define $\theta=\theta_C$ as a virtual control input for the elevation dynamics
\begin{equation}\label{eq: Tangential Dynamics}
\ddot{\alpha}=-\frac{1}{r}\left(2\dot{r}\dot{\alpha}+g\cos\alpha\right)+\frac{1}{mr}u_{1}\cos\left(\alpha+\theta_C\right)
\end{equation}
subject to the cable tension
\[
T=mr\dot{\alpha}^{2}-mg\sin\alpha+u_{1}\sin\left(\alpha+\theta_C\right)-m\ddot{r}.
\]
The goal of the outer loop is to satisfy simultaneously conditions (\ref{eq: Control Objective1})-(\ref{eq: Control Objective2})
using the modulus and direction of the thrust vector. The following proposition provides a suitable control law.
\begin{prop}
Let systems (\ref{eq: Radial Dynamics}), (\ref{eq: Tangential Dynamics}) be subject to constraint
(\ref{eq: Taut Cable Constraint}) where $\theta=\theta_C$ is a control input. Let (\ref{eq: u3}) be the control law for the radial dynamics (\ref{eq: Radial Dynamics}) and let (\ref{eq: Tangential Dynamics}) be controlled by
\begin{eqnarray}
&&u_{1}=\sqrt{u_{T}^{2}+u_{\alpha}^{2}} \label{eq: u1}\\
&&\theta_C=\frac{\pi}{2}-\alpha-\mathrm{atan2}\left(u_{\alpha},u_{T}\right) \label{eq: thetaC}
\end{eqnarray}
with
\begin{eqnarray}
&& u_{T}=\bar{T}+mg\sin\alpha+m\ddot{r}\label{eq: uT}\\
&& u_{\alpha}=m\left(2\dot{r}\dot{\alpha}+g\cos\alpha\right)\!-\!mr\left(k_{P\alpha}\!\left(\alpha\!-\!\bar{\alpha}\right)\!+\!k_{D\alpha}\dot{\alpha}\right)\label{eq: uA}
\end{eqnarray}
with $\bar{T}$ as in (\ref{eq: Equilibrium Tension}). Given the reference $\left[\bar{r},\bar{\alpha},\bar{\theta}\right]\in\mathcal{S}_{\epsilon}$, the control objectives in Problem \ref{prb: ControlObj} are satisfied for $k_{P\alpha}>0$, $k_{D\alpha}>0$ and $\lambda_{1}<\frac{\bar{T}}{m}$.
\end{prop}
\begin{pf}
Following from Proposition \ref{prop: Radial Control}, the radial dynamics (\ref{eq: Radial Dynamics}) asymptotically tend to $\bar{r}$ and $\left\Vert \ddot{r}\right\Vert _{\infty}\leq\lambda_{1}$. As for the elevation dynamics (\ref{eq: Tangential Dynamics}) and
the cable constraint (\ref{eq: Taut Cable Constraint}), by substituting
\begin{equation}
u_{1}\left[\begin{array}{c}
\cos\left(\alpha+\theta_C\right)\\
\sin\left(\alpha+\theta_C\right)
\end{array}\right]=\left[\begin{array}{c}
u_{\alpha}\\
u_{T}
\end{array}\right].\label{eq: VecThrust}
\end{equation}
it follows that
\[
\left\{ \begin{array}{l}
\ddot{\alpha}=-\frac{1}{r}\left(2\dot{r}\dot{\alpha}+g\cos\alpha\right)+\frac{1}{mr}u_{\alpha}\\
T=mr\dot{\alpha}^{2}-mg\sin\alpha-m\ddot{r}+u_{T}.
\end{array}\right.
\]
Then, given (\ref{eq: uT}) and (\ref{eq: uA}), system (\ref{eq: Tangential Dynamics}) becomes
\[
\left\{ \begin{array}{l}
\ddot{\alpha}=-k_{P\alpha}\left(\alpha-\bar{\alpha}\right)-k_{D\alpha}\dot{\alpha}\\
T=\bar{T}+mr\dot{\alpha}^{2}
\end{array}\right.
\]
which satisfies conditions (\ref{eq: Control Objective1})-(\ref{eq: Control Objective2}). As a result, system \eqref{eq: Taut Cable Dynamics} asymptotically
tends to $\left[\bar{r},\bar{\alpha},\bar{\theta}\right]$ without
violating the taut cable condition. Equations \eqref{eq: u1} and \eqref{eq: uA} follow directly from (\ref{eq: VecThrust}).

To conclude the proof, note that equation (\ref{eq: thetaC}) is undefined if $u_{T}=u_{\alpha}=0$.
However, this condition is never verified since $u_{T}\geq\bar{T}-m\left\Vert \ddot{r}\right\Vert _{\infty}>0$
due to the conditions $\lambda_{1}<\frac{\bar{T}}{m}$ and $\alpha\in\left[-\pi,\pi\right]$.
$\hfill\Box$ \end{pf}

The proposed outer loop satisfies all the control objectives under the assumption that the inner loop is ideal.
However, the following section will show how the presence of a real inner loop can cause a degradation of the outer loop performances.

\subsection{Inner Loop Control}

The presence of an attitude error
\begin{equation}\label{eq: AttitudeErrorCoordinates}
\tilde{\theta} :=\theta-\theta_{C}
\end{equation}
has the double effect of modifying the UAV tangential dynamics as
well as the cable tension. Indeed, by substituting $\theta=\theta_{C}+\tilde{\theta}$
into equations (\ref{eq: Tension})-(\ref{eq: Taut Cable Dynamics}), the following expressions are obtained:
\begin{equation}
\begin{array}{l}
\!\!\!\!\!\!\ddot{{\alpha}}\!=\!-\displaystyle\frac{1}{r}\left(2\dot{r}\dot{{\alpha}}\! + \!g\cos(\tilde{\alpha}\!+\! \bar{\alpha}) \right)\! +\!
\displaystyle\frac{1}{mr}u_{1}\cos\left(\alpha\!+\!\theta_C\!+\!\tilde\theta\right)\\
\!\!\!\!\!\!T\!=\!mr\dot{{\alpha}}^{2}\!-\!mg\sin (\tilde{\alpha} \!+\! \bar{\alpha}) \!+\! u_{1}\sin\left(\!\alpha\!+\!\theta_C\!+\!\tilde\theta\!\right)\!-\!m\ddot{r}
\end{array}\label{eq: Influenza di ThetaTilde}
\end{equation}
where $\tilde{\alpha}$ is the elevation angle error
\begin{equation}\label{eq: AlphaErrorCoordinates}
\tilde{\alpha}:=\alpha-\bar{\alpha}\,.
\end{equation}
By substituting
\[
\begin{array}{c}
\cos\left(\alpha+\theta_C+\tilde\theta\right)=\cos\left(\alpha+\theta_C\right)\cos\tilde{\theta}-\sin\left(\alpha+\theta_C\right)\sin\tilde{\theta}\\
\sin\left(\alpha+\theta_C+\tilde\theta\right)=\sin\left(\alpha+\theta_C\right)\cos\tilde{\theta}+\cos\left(\alpha+\theta_C\right)\sin\tilde{\theta}
\end{array}
\]
and taking into account (\ref{eq: uT})-(\ref{eq: VecThrust}),
equations (\ref{eq: Influenza di ThetaTilde}) become
\begin{eqnarray}
&&\ddot{{\alpha}}\!=\!-\!\left(k_{P\alpha}\tilde{\alpha}\!+\! k_{D\alpha}\!\dot{\tilde{\alpha}}\right)\!\cos\tilde{\theta}\!+\!\Delta\!\left(\!\frac{\dot{r}}{r},\tilde{\theta}\!\right)\!\dot{\tilde{\alpha}}\!+\!\Gamma\!\left(\! r,\ddot{r},\tilde{\alpha},\tilde{\theta}\!\right)\!\!\label{eq: TanDyn_Real}\\
&&T=mr\dot{\tilde{\alpha}}^{2}+\bar{T}-u_{T}\!\left(\!1-\cos\tilde{\theta}\!\right)+u_{\alpha}\sin\tilde{\theta}\label{eq: Treal}
\end{eqnarray}
with
\[
\begin{array}{l}
\Delta\left(\frac{\dot{r}}{r},\tilde{\theta}\right)=\frac{2\dot{r}}{r}\left(1-\cos\tilde{\theta}\right)\\
\Gamma\left(r,\ddot{r},\tilde{\alpha},\tilde{\theta}\right)=\frac{g\cos(\tilde{\alpha}+\bar{\alpha})}{r}\left(\cos\tilde{\theta}-1\right)-\frac{1}{mr}u_{T}\sin\tilde{\theta}.
\end{array}
\]
In summary, in the presence of the attitude error $\tilde{\theta}$, the outer loop behaviour is given by (\ref{eq: TanDyn_Real}), which is a system with state $x_\alpha := [\tilde{\alpha},\,\dot{\alpha}]^T$ and affected by the exogenous inputs $\tilde\theta$, $\dot{r} / r$, and $\Gamma(\cdot)$. For this system the following result holds true.

\begin{prop}
\label{prop: IOS Outer Loop}
System (\ref{eq: TanDyn_Real}) is ISS with no restrictions on the initial conditions, no restriction on the input $\Gamma(\cdot)$ and with restriction $|\tilde\theta|\leq\tilde\theta_{\max}$ and restriction
 \begin{equation}
\left\vert\frac{r}{\dot{r}}\right\vert\leq R=\nu \frac{k_{D\alpha}}{2}\frac{\cos\tilde{\theta}_{\max}}{1-\cos\tilde{\theta}_{\max}},\label{eq: IOS restrictions OutLoop}
\end{equation}
where $\tilde\theta_{\max}\in(0,\pi/2)$ and $\nu\in(0,1)$. Moreover, there exists a finite asymptotic gain $\gamma_\mathrm{Out}$ between the disturbance $\tilde{\theta}$ and the output $y_\alpha:= \dot{\theta}_C$.
\end{prop}
\begin{pf}
The Proof is provided in Appendix B.
$\hfill\Box$ \end{pf}

Let us now focus on the inner attitude loop which is given by the third equation in (\ref{eq: Taut Cable Dynamics}). By choosing the control input $u_2$ as
\begin{equation} \label{eq: u2}
u_{2}=-\mathcal{J}\left( k_{P\theta}\tilde{\theta}+k_{D\theta}\dot{\tilde{\theta}}\right),
\end{equation}
where $k_{P\theta},\, k_{D\theta} \in \Real_{> 0}$ are control parameters to be tuned,
the attitude error dynamics become

\begin{equation}
\left\{ \begin{array}{ll}
\dot{\tilde{\theta}} &= \dot{\theta} - \dot{\theta}_C\\
\ddot{\theta} &= - k_{P\theta}\tilde{\theta} - k_{D\theta} \dot{\theta}\,.
\end{array}\right.\label{eq: Inner Loop Error}
\end{equation}

For system (\ref{eq: Inner Loop Error}), which is a system with state $x_{\theta} := [ \tilde{\theta},\, \dot{\theta}]^T$ affected by the exogenous input $\dot{\theta}_C$, the following result holds true.

\begin{prop}
\label{prop: IOS Inner Loop}
Consider the closed loop system (\ref{eq: Inner Loop Error}).
Let $k_{P\theta}$ and $k_{D\theta}$ be chosen as $k_{D\theta}=2\zeta\sqrt{k_{P\theta}}$, $k_{P\theta}>0$ with $\zeta\in\left(0,1\right)$. Then the following results hold true:
\begin{itemize}
\item system (\ref{eq: Inner Loop Error}) is ISS with respect to the reference velocity $\dot{\theta}_{C}$;
\item given $k_{P\theta}>1$, the asymptotic gain $\gamma_{\mathrm{In}}$ between the disturbance $\dot{\theta}_{C}$
and the output $y_{\theta} :=\tilde{\theta}$ satisfies
\[
\gamma_{\mathrm{In}} \leq  \frac{1}{\zeta\sqrt{k_{P\theta}}}.
\]
\end{itemize}
\end{prop}
\begin{pf}
See Appendix C.
$\hfill\Box$ \end{pf}

With Propositions \ref{prop: Radial Control}, \ref{prop: IOS Outer Loop} and \ref{prop: IOS Inner Loop} at hand, it is now possible to derive the main stability results pertaining to the overall interconnected system.

\begin{prop}\label{prop: Small Gain}
Let system (\ref{eq: Taut Cable Dynamics}) be subject to control inputs (\ref{eq: u3}), (\ref{eq: u1}) and (\ref{eq: u2}). Given a sufficiently high inner loop gain $k_{P\theta}$ and a bounded saturation value
\begin{equation}\label{eq: Lambda1}
\lambda_{1}<\nu\frac{k_{D\alpha}k_{Dr}r_{\min}}{2}\frac{\cos\left(\tilde\theta_{\max}\right)}{1-\cos\left(\tilde\theta_{\max}\right)},
\end{equation}
with $\nu\in(0,1)$ and $\tilde\theta_{\max}\in(0,\frac{\pi}{2})$, the setpoint $\left[\bar{r},\bar{\alpha},\bar{\theta}\right]\in\mathcal{S}_{\epsilon}$ is asymptotically stable for any initial conditions satisfying
\begin{equation}\label{eq: Init Rdot}
\left|\dot{r}\left(0\right)\right|\leq\frac{\lambda_{1}}{k_{Dr}}
\end{equation}
and
\begin{equation}\label{eq: Init Conditions}
\left\Vert x_{\mathrm{\theta}}\left(0\right)\right\Vert +\gamma_{\mathrm{In}}\left\Vert x_{\mathrm{\alpha}}\left(0\right)\right\Vert <\left(1-\gamma_{\mathrm{In}}\gamma_{\mathrm{Out}}\right)\left(\tilde\theta_{\max}\right).
\end{equation}
Moreover, the following bound holds true
\begin{equation}\label{eq: InfNorm Bound}
\left[\begin{array}{c}
\left\Vert x_{\alpha}\right\Vert _{\infty}\\
\left\Vert x_{\theta}\right\Vert _{\infty}
\end{array}\right]\leq\frac{1}{1-\gamma_{\mathrm{In}}\gamma_{\mathrm{Out}}}\left[\begin{array}{cc}
1 & \gamma_{\mathrm{Out}}\\
\gamma_{\mathrm{In}} & 1
\end{array}\right]\left[\begin{array}{c}
\left\Vert x_{\alpha}\left(0\right)\right\Vert \\
\left\Vert x_{\theta}\left(0\right)\right\Vert
\end{array}\right].
\end{equation}
\end{prop}

\begin{pf}
As proven in Proposition \ref{prop: Radial Control}, the radial dynamics are such that $r>0$ and $r(t)$ asymptotically tends to $\bar{r}$ independently from the rest of system.

Following from Proposition \ref{prop: IOS Outer Loop}, subsystem (\ref{eq: TanDyn_Real}) is ISS with an asymptotic gain $\left\Vert\dot\theta_C\right\Vert\leq\gamma_{Out}\left\Vert\tilde\theta\right\Vert$ if
\begin{itemize}
\item $\left|\dot{r}/r\right|\leq R$ with $R$ as in (\ref{eq: IOS restrictions OutLoop}). This restriction is always satisfied due to conditions (\ref{eq: Lambda1}) and (\ref{eq: Init Rdot}).
\item $\left\vert \tilde{\theta}(t)\right\vert\leq\tilde{\theta}_{\max}$ with $\tilde{\theta}_{\max}\in\left(0,\frac{\pi}{2}\right)$.
\end{itemize}
Following from Proposition \ref{prop: IOS Inner Loop}, subsystem (\ref{eq: Inner Loop Error}) is ISS with an asymptotic gain
\[
\left\Vert\tilde\theta\right\Vert\leq\frac{\left\Vert\dot\theta_C\right\Vert}{\zeta\sqrt{k_{P\theta}}}.
\]
Therefore, given
\[
k_{P\theta}\geq\frac{\gamma_{Out}^2}{\zeta^2},
\]
the small gain condition for the interconnected systems is satisfied at least at time $t=0$. As long as the small gain theorem is applicable, the trajectories of the interconnected systems are bounded by (\ref{eq: InfNorm Bound}).
Therefore, by choosing initial conditions such that (\ref{eq: Init Conditions}) holds true, it follows that $\left\Vert \tilde{\theta}\right\Vert_{\infty}\leq\tilde{\theta}_{\max}$ and therefore the small gain theorem remains applicable at all times.
$\hfill\Box$ \end{pf}

The main interest of Proposition \ref{prop: Small Gain} is that it not only proves the Asymptotic Stability of the desired set-point, but it also provides an explicit bound for
the system trajectories. This implies that the state trajectories of the system are limited for any set-point of the closed-loop system. Starting from the stabilized
system obtained in this section, the following section will provide a strategy that systematically changes the reference of the closed-loop system so that the constraints are
satisfied at all times.

\section{REFERENCE GOVERNOR}\label{sec: REFERENCE GOVERNOR}

\begin{figure}
\includegraphics[scale=0.7]{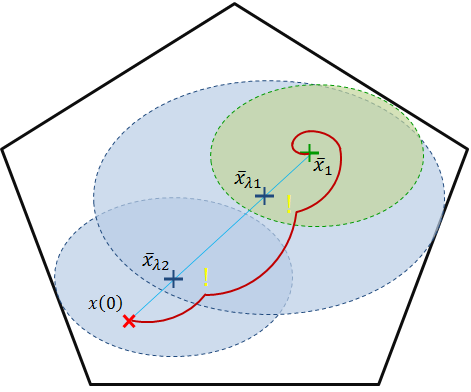}\caption{\label{fig: RGIdea}Basic idea of the proposed Reference Governor.
The exclamation marks denote the instant at which the applied reference
is changes from $\bar{x}_{\lambda2}$ to $\bar{x}_{\lambda1}$ and
from $\bar{x}_{\lambda1}$ to $\bar{x}_{1}$.}
\end{figure}

This section will develop an ad-hoc Reference Governor that, whenever necessary, modifies
the desired reference $\left[\bar{r},\bar{\alpha},\bar{\theta}\right]\in\mathcal{S}_\epsilon$
into a succession of intermediate waypoints
$\left[\bar{r}_{k},\bar{\alpha}_{k},\bar{\theta}_{k}\right]\in\mathcal{S}_\epsilon$
to prevent the violation of constraints. The main idea follows from the results of
Proposition \ref{prop: Small Gain}: since bounded initial conditions imply
bounded trajectories, any attainable equilibrium point is characterized
by a set of initial conditions that do not violate the system constraints.
The idea is to steer the system from one waypoint to
the next until the desired setpoint is applicable without violating
the constraints. The basic idea is depicted in Figure \ref{fig: RGIdea}.
The following definition is given.
\begin{defn}
\label{def. Ball of initial conditions}Given a generic reference
$\left[\bar{r}_{k},\bar{\alpha}_{k},\bar{\theta}_{k}\right]$,
the set $\mathcal{I}_{k}$ of suitable initial conditions is
defined such that the closed-loop system verifies
\[
\left[\!\!\!\begin{array}{c}
x_{r}\left(0\right)\\
x_{\alpha}\left(0\right)\\
x_{\theta}\left(0\right)
\end{array}\!\!\!\right]\!\!\in\!\mathcal{I}_{k}\Rightarrow\left\{ \begin{array}{l}
\underset{t\rightarrow\infty}{\lim}\left[r\left(t\right),\alpha\left(t\right),\theta\left(t\right)\right]=\left[\bar{r}_{k},\bar{\alpha}_{k},\bar{\theta}_{k}\right]\\
T\left(r\left(t\right),\alpha\left(t\right),\theta\left(t\right)\right)>0\quad\forall t\in\left[0,\infty\right).
\end{array}\right.
\]
\end{defn}
In the absence of the Reference Governor, Definition \ref{def. Ball of initial conditions}
implies that the system is guaranteed to converge to the setpoint
$\left[\bar{r}_{0},\bar{\alpha}_{0},\bar{\theta}_{0}\right]:=\left[\bar{r},\bar{\alpha},\bar{\theta}\right]$ without
violating the constraints if and only if $\left[x_{r}\left(0\right);\,x_{\alpha}\left(0\right);\,x_{\theta}\left(0\right)\right]\in\mathcal{I}_{0}$.
The objective of the RG is to extend the set of initial conditions that can be led to the desired reference without violating the constraints.
To do so, consider a waypoint $\left[\bar{r}_{1},\bar{\alpha}_{1},\bar{\theta}_{1}\right]\in\mathcal{S}_{\epsilon}$ such that
\begin{equation}\label{eq: RG WaypointCondition}
\left[
\bar{r}_{k+1}-\bar{r}_k,\:
0,\:
\bar{\alpha}_{k+1}-\bar{\alpha}_k,\:
0,\:
\bar{\theta}_{k+1}-\bar{\theta}_k,\:
0
\right]^T\!\!\in\mathcal{I}_k
\end{equation}
and
\begin{equation}\label{eq: RG SetGrowth}
\mathcal{I}_{k+1}\setminus\mathcal{I}_{k}\neq\left\{ \textrm{{\O}}\right\},
\end{equation}
with $k=0$ for the time being. Given an initial condition $\left[x_{r}\left(0\right);\,x_{\alpha}\left(0\right);\,x_{\theta}\left(0\right)\right]\in\mathcal{I}_{1}$, it is possible to guarantee constraint satisfaction by providing $\left[\bar{r}_{1},\bar{\alpha}_{1},\bar{\theta}_{1}\right]$
as a temporary reference. Since the closed-loop system asymptotically tends to the waypoint $\left[\bar{r}_{1},\bar{\alpha}_{1},\bar{\theta}_{1}\right]$, it follows from condition \eqref{eq: RG WaypointCondition} that there exists a finite
time $\tau$ after which
\begin{equation}
\left\{ \begin{array}{l}
\left[x_{r}\left(\tau\right);\,x_{\alpha}\left(\tau\right);\,x_{\theta}\left(\tau\right)\right]\in\mathcal{I}_{0}\\
T\left(r\left(t\right),\alpha\left(t\right),\theta\left(t\right)\right)>0\quad\forall t\in\left[0,\tau\right],
\end{array}\right.\label{eq: RG Objective}
\end{equation}
is verified. As a result, by changing the reference to $\left[\bar{r}_{0},\bar{\alpha}_{0},\bar{\theta}_{0}\right]$ at $t=\tau$, the introduction of the intermediate waypoint $\left[\bar{r}_{1},\bar{\alpha}_{1},\bar{\theta}_{1}\right]$ can be used to reach the final setpoint from any initial condition belonging to the set $\mathcal{I}_{1}\cup\mathcal{I}_{0}$. This set is strictly larger than $\mathcal{I}_{0}$ due to condition \eqref{eq: RG SetGrowth}. By applying the algorithm recursively, it follows that the final setpoint can be attained without violating the constraints if
\begin{equation}
\left[x_{r}\left(0\right),x_{\alpha}\left(0\right),x_{\theta}\left(0\right)\right]\in\mathcal{I}_{0}\cup\mathcal{I}_{1}\cup\mathcal{I}_{2}\cup\ldots\cup\mathcal{I}_{K}\label{eq: SetSumObjective}
\end{equation}
where $\left[\bar{r}_{K},\bar{\alpha}_{K},\bar{\theta}_{K}\right]\in\mathcal{S}_{\epsilon}$
is an arbitrary starting point chosen such that $\left[x_{r}\left(0\right),x_{\alpha}\left(0\right),x_{\theta}\left(0\right)\right]\in\mathcal{I}_{K}$.
\begin{rem}
For the sake of simplicity, this paper only addresses the case $\left[r\left(0\right),\alpha\left(0\right),\theta\left(0\right)\right]\in\mathcal{S}_{\epsilon}$
which enables the choice $\left[\bar{r}_{K},\bar{\alpha}_{K},\bar{\theta}_{K}\right]=\left[r\left(0\right),\alpha\left(0\right),\theta\left(0\right)\right]$ with a
limitation on the maximum starting velocities. However, the set of admissible initial conditions can potentially be extended to $\left[x_{r}\left(0\right),x_{\alpha}\left(0\right),x_{\theta}\left(0\right)\right]\in\mathcal{I}_{\mathcal{S}\epsilon}$,
where $\mathcal{I}_{\mathcal{S}\epsilon}$ is given by the union of
all the $\mathcal{I}_{\lambda}$ belonging to the set $\mathcal{S}_{\epsilon}$. Please note that the control performances of the backtracking reference governor will be only marginally affected by the choice of $\left[\bar{r}_{K},\bar{\alpha}_{K},\bar{\theta}_{K}\right]$.
\end{rem}
Having defined the main strategy of the Reference Governor, it follows
that the necessary steps for its development are:
\begin{enumerate}
\item Given any reference $\left[\bar{r}_{k},\bar{\alpha}_{k},\bar{\theta}_{k}\right]\in\mathcal{S}_{\epsilon}$,
define a set of suitable initial conditions which cannot lead to constraint violation.
\item Show that, for any two references $\left[\bar{r}_{K},\bar{\alpha}_{K},\bar{\theta}_{K}\right]$
and $\left[\bar{r}_{0},\bar{\alpha}_{0},\bar{\theta}_{0}\right]$
belonging to $\mathcal{S}_{\epsilon}$, it is possible to provide
a continuous curve of attainable equilibrium points $\left[\bar{r}_{\lambda},\bar{\alpha}_{\lambda},\bar{\theta}_{\lambda}\right]\in\mathcal{S}_{\epsilon}$
connecting these two references.
\item Provide an algorithm for calculating a succession of waypoints $\left[\bar{r}_{k},\bar{\alpha}_{k},\bar{\theta}_{k}\right]\in\left[\bar{r}_{\lambda},\bar{\alpha}_{\lambda},\bar{\theta}_{\lambda}\right]$
such that each waypoint satisfies conditions \eqref{eq: RG WaypointCondition}-\eqref{eq: RG SetGrowth}.
\item Determine the conditions for switching the applied reference from
the current waypoint to the next one.
\end{enumerate}
It is worth noting that although the first two steps are specific
to the system at hand, the method can be generalized to any
closed-loop nonlinear system subject to constraints.\\

The following proposition addresses the first step of the RG by analytically providing an inner approximation of the set $\mathcal{I}_k$ associated to $\left[\bar{r}_{k},\bar{\alpha}_{k},\bar{\theta}_{k}\right]\in\mathcal{S}_{\epsilon}$.

\begin{prop}\label{prop: Ball of initial conditions}
For any reference $\left[\bar{r}_{k},\bar{\alpha}_{k},\bar{\theta}_{k}\right]\in\mathcal{S}_{\epsilon}$, there exists a set of initial conditions $r(0)>\frac{\lambda_1}{k^2_{Dr}}, \left|\dot{r}\left(0\right)\right|\leq\frac{\lambda_{1}}{k_{Dr}}$ and
\[
\begin{array}{c}
\left|x_{\alpha}\left(0\right)\right|\leq\Delta x_{\alpha k}\\
\left|x_{\theta}\left(0\right)\right|\leq\Delta x_{\theta k}
\end{array}
\]
such that $T\left(t\right)>0\;\forall t\in\left[0,\infty\right)$.
Moreover, there exist two positive constants $\delta_{\alpha},\delta_{\theta}>0$ such that $\Delta x_{\alpha k}\geq\delta_{\alpha}$ and $\Delta x_{\theta k}\geq\delta_{\theta}$.
\end{prop}
\begin{pf}
Following from expression (\ref{eq: Treal}), the taut cable constraint
is satisfied if
\[
\left\Vert u_{T}\left(1-\cos\tilde{\theta}\right)\right\Vert +\left\Vert u_{\alpha}\sin\tilde{\theta}\right\Vert <\bar{T}_{k}.
\]
Referring to equations (\ref{eq: uT}) and (\ref{eq: uA}), this condition
can be bounded by
\[
\begin{array}{l}
\left(\bar{T}_{k}+\sqrt{2}mg+m\left\Vert u_{3}\right\Vert _{\infty}\right)\left\Vert x_{\theta}\right\Vert _{\infty}+\\
\quad\left(2m\left\Vert \dot{r}\right\Vert _{\infty}+m\left\Vert r\right\Vert _{\infty}\left(k_{P\alpha}+k_{D\alpha}\right)\right)\left\Vert x_{\theta}\right\Vert _{\infty}\left\Vert x_{\alpha}\right\Vert _{\infty}<\bar{T}_{k}.
\end{array}
\]
To satisfy the inequality, it is sufficient to limit the infinity
norms
\[
\left\Vert x_{\theta}\right\Vert _{\infty}<\frac{\bar{T}_{k}}{\bar{T}_{k}+\sqrt{2}mg+m\left\Vert u_{3}\right\Vert _{\infty}}
\]
and
\[
\left\Vert x_{\alpha}\right\Vert _{\infty}<\frac{\bar{T}_{k}-\left(\bar{T}_{k}+\sqrt{2}mg+m\left\Vert u_{3}\right\Vert _{\infty}\right)\left\Vert x_{\theta}\right\Vert _{\infty}}{\left(2m\left\Vert \dot{r}\right\Vert _{\infty}+m\left\Vert r\right\Vert _{\infty}\left(k_{P\alpha}+k_{D\alpha}\right)\right)\left\Vert x_{\theta}\right\Vert _{\infty}}.
\]
Following from Proposition \ref{prop: Small Gain}, the infinity norms are
bounded by the initial conditions via expression (\ref{eq: InfNorm Bound}).
Therefore, by choosing
\[
\left[\begin{array}{c}
\Delta x_{\theta k}\\
\Delta x_{\alpha k}
\end{array}\right]\leq\left(1-\gamma_{\mathrm{In}}\gamma_{\mathrm{Out}}\right)\left[\begin{array}{cc}
1 & \gamma_{\mathrm{In}}\\
\gamma_{\mathrm{Out}} & 1
\end{array}\right]^{-1}\left[\begin{array}{c}
\left\Vert x_{\theta}\right\Vert _{\infty}\\
\left\Vert x_{\alpha}\right\Vert _{\infty}
\end{array}\right].
\]
The taut cable constraint is satisfied for $\left|x_{\alpha}\left(0\right)\right|\leq\Delta x_{\alpha k}$
and $\left|x_{\theta}\left(0\right)\right|\leq\Delta x_{\theta k}$.
Moreover, by taking $\bar{T}_{k}=\epsilon$ and calculating
the corresponding $\delta_{\theta}$ and $\delta_{\alpha}$, it follows
that $\Delta x_{\theta k}\geq\delta_{\theta}$ and $\Delta x_{\alpha}\geq\delta_{\alpha}$
regardless of the equilibrium point.
$\hfill\Box$ \end{pf}
Having shown that any attainable reference is characterized by a
set of suitable initial conditions, the second step of the RG is to
define a continuous curve contained in $\mathcal{S}_{\epsilon}$ and
connecting any two attainable references. In view of using a linear
interpolation to define such curve, the following proposition shows
that the set $\mathcal{S}_{\epsilon}$ can be divided into two convex
sets overlapping in $\bar{\alpha}=\frac{\pi}{2}$.
\begin{prop}
\label{prop: Curve of Eq Points}Given the final reference $\left[\bar{r}_{0},\bar{\alpha}_{0},\bar{\theta}_{0}\right]\in\mathcal{S}_{\epsilon}$
and the initial reference $\left[\bar{r}_{K},\bar{\alpha}_{K},\bar{\theta}_{K}\right]\in\mathcal{S}_{\epsilon}$,
if $\bar{\alpha}_{0},\bar{\alpha}_{K}$ both belong to the same interval
\textup{$\left[0,\frac{\pi}{2}\right]$ }or \textup{$\left[\frac{\pi}{2},\pi\right]$},
then, the curve \textup{
\begin{equation}
\left[\bar{r}_{\lambda},\bar{\alpha}_{\lambda},\bar{\theta}_{\lambda}\right]=\lambda\left[\bar{r}_{0},\bar{\alpha}_{0},\bar{\theta}_{0}\right]+\left(1-\lambda\right)\left[\bar{r}_{K},\bar{\alpha}_{K},\bar{\theta}_{K}\right]\label{eq: WaypointDef}
\end{equation}
}belongs to the set \textup{$\mathcal{S}_{\epsilon}\;\forall\lambda\in\left[0,1\right]$.}\end{prop}
\begin{pf}
Consider the case $\bar{\alpha}_{K}\in\left[0,\frac{\pi}{2}\right]$
and $\bar{\alpha}_{0}\in\left[0,\frac{\pi}{2}\right]$, it follows
that
\[
\bar{\alpha}_{\lambda}=\lambda\bar{\alpha}_{0}+\left(1-\lambda\right)\bar{\alpha}_{K}\in\left[0,\frac{\pi}{2}\right].
\]
For $\bar{\alpha}_{\lambda}\in\left[0,\frac{\pi}{2}\right]$, the
equilibrium point belongs to $\mathcal{S}_{\epsilon}$ if
\[
\bar{\theta}_{\lambda}\leq\mathrm{arcot}\left(\frac{\epsilon}{mg\cos\bar{\alpha}_{\lambda}}+\tan\bar{\alpha}_{\lambda}\right)
\]
which can be rewritten as
\begin{equation}
\bar{\theta}_{\lambda}\leq\varsigma\left(\bar{\alpha}\right)\label{eq: Concave function}
\end{equation}
where
\[
\varsigma\left(\bar{\alpha}\right)=\mathrm{arcot}\left(\frac{\epsilon}{mg\cos\bar{\alpha}_{\lambda}}+\tan\bar{\alpha}_{\lambda}\right)
\]
is a concave function since
\[
\frac{\partial^{2}\varsigma}{\partial\bar{\alpha}^{2}}=-\frac{d\left(d^{2}-1\right)\cos\bar{\alpha}_{\lambda}}{\left(2d\sin\bar{\alpha}_{\lambda}+d^{2}+1\right)^{2}}\leq0.
\]
As a result, inequality (\ref{eq: Concave function}) is a convex
constraint and the choice $\bar{\theta}_{\lambda}=\lambda\bar{\theta}_{0}+\left(1-\lambda\right)\bar{\theta}_{K}$
leads to
\[
\left[\bar{r}_{\lambda},\bar{\alpha}_{\lambda},\bar{\theta}_{\lambda}\right]\in\mathcal{S}_{\epsilon}.
\]
The case $\bar{\alpha}_{0}\in\left[\frac{\pi}{2},\pi\right]$ and
$\bar{\alpha}_{\lambda0}\in\left[\frac{\pi}{2},\pi\right]$ can be proven
analogously.
$\hfill\Box$ \end{pf}
As a result, if $\bar{\alpha}_{0},\bar{\alpha}_{K}$ belong to the
same interval, the curve of attainable waypoints is generated using
linear interpolation. If $\bar{\alpha}_{0},\bar{\alpha}_{K}$ do not
belong to the same interval, the curve of attainable waypoints can be
generated using a piecewise linear chain connecting the starting reference
$\left[\bar{r}_{K},\bar{\alpha}_{K},\bar{\theta}_{K}\right]$ to the
overlap point $\left[\frac{\bar{r}_{0}+\bar{r}_{K}}{2},\frac{\pi}{2},0\right]$
and then proceeding from $\left[\frac{\bar{r}_{0}+\bar{r}_{K}}{2},\frac{\pi}{2},0\right]$
to the final destination $\left[\bar{r}_{0},\bar{\alpha}_{0},\bar{\theta}_{0}\right]$.
The third step of the proposed RG strategy is to define a suitable
succession of waypoints to use as intermediate references. This paper
introduces a backtracking algorithm that iteratively defines waypoints
in such a way that (\ref{eq: SetSumObjective}) is verified.

\subsection{Backtracking Algorithm}

Given the final reference $\left[\bar{r}_{0},\bar{\alpha}_{0},\bar{\theta}_{0}\right]\in\mathcal{S}_{\epsilon}$
and the starting reference $\left[\bar{r}_{K},\bar{\alpha}_{K},\bar{\theta}_{K}\right]\in\mathcal{S}_{\epsilon}$,
the backtracking algorithm is charged with defining the succession
of waypoints $\left[\bar{r}_{k},\bar{\alpha}_{k},\bar{\theta}_{k}\right]\in\mathcal{S}_{\epsilon}$
such that, for $k=1,...,K$, conditions \eqref{eq: RG WaypointCondition}-\eqref{eq: RG SetGrowth} are respected.
To do so, consider the final reference $\left[\bar{r}_{0},\bar{\alpha}_{0},\bar{\theta}_{0}\right]\in\mathcal{S}_{\epsilon}$
and starting waypoint $\left[\bar{r}_{K},\bar{\alpha}_{K},\bar{\theta}_{K}\right]\in\mathcal{S}_{\epsilon}$
such that $\bar{\alpha}_{0},\bar{\alpha}_{K}$ belong to the same
interval $\left[0,\frac{\pi}{2}\right]$ or $\left[\frac{\pi}{2},\pi\right]$.
Following from Proposition \ref{prop: Curve of Eq Points}, any point
belonging to the segment
\[
\left[\bar{r}_{\lambda},\bar{\alpha}_{\lambda},\bar{\theta}_{\lambda}\right]=\lambda\left[\bar{r}_{\lambda0},\bar{\alpha}_{\lambda0},\bar{\theta}_{\lambda0}\right]+\left(1-\lambda\right)\left[\bar{r}_{0},\bar{\alpha}_{0},\bar{\theta}_{0}\right]
\]
is an attainable equilibrium point. The only question is how to choose a suitable value for
the parameter $\lambda_{1}\in\left(0,1\right]$. By taking advantage of Proposition \ref{prop: Ball of initial conditions},
it follows that conditions \eqref{eq: RG WaypointCondition}-\eqref{eq: RG SetGrowth} are both satisfied if
\[
\begin{array}{c}
\left|\bar{\alpha}_{0}-\bar{\alpha}_{1}\right|=\Delta x_{\alpha1}-\frac{\delta_{\alpha}}{2}\\
\left|\bar{\theta}_{0}-\bar{\theta}_{1}\right|=\Delta x_{\theta1}-\frac{\delta_{\theta}}{2}.
\end{array}
\]
As a result, the last waypoint can be chosen as
\[
\lambda_{1}=\max\left(0,\;1-\frac{\Delta x_{\alpha1}-\frac{\delta_{\alpha}}{2}}{\left|\bar{\alpha}_{0}-\bar{\alpha}_{K}\right|},\;1-\frac{\Delta x_{\theta1}-\frac{\delta_{\theta}}{2}}{\left|\bar{\theta}_{0}-\bar{\theta}_{K}\right|}\right).
\]
Given the last waypoint, the second to last waypoint (and the following
ones) can be calculated iteratively using
\[
\lambda_{k+1}=\max\left(0,\;1-\frac{\Delta x_{\alpha}\left(\lambda_{k}\right)-\frac{\delta_{\alpha}}{2}}{\left|\bar{\alpha}_{k}-\alpha\left(0\right)\right|},\;1-\frac{\Delta x_{\theta}\left(\lambda_{k}\right)-\frac{\delta_{\theta}}{2}}{\left|\bar{\theta}_{k}-\theta\left(0\right)\right|}\right).
\]
The process is terminated when $\lambda_{k+1}=0$ which implies
$\left[x_r(0);x_\alpha(0);x_\theta(0)\right]\in\mathcal{I}_{K}$.\\

If $\bar{\alpha}_{0},\bar{\alpha}_{K}$ do not belong to the same
interval $\left[0,\frac{\pi}{2}\right]$ or $\left[\frac{\pi}{2},\pi\right]$,
the backtracking algorithm must be applied twice: first to define
the succession of waypoints connecting $\left[\bar{r}_{0},\bar{\alpha}_{0},\bar{\theta}_{0}\right]$
to $\left[\frac{\bar{r}_{0}+\bar{r}_{K}}{2},\frac{\pi}{2},0\right]$,
then to define the succession between $\left[\frac{\bar{r}_{0}+\bar{r}_{K}}{2},\frac{\pi}{2},0\right]$
and $\left[\bar{r}_{K},\bar{\alpha}_{K},\bar{\theta}_{K}\right]$.

\subsection{Switching Conditions}

The final thing left to consider is when should the reference governor
change the reference from one waypoint to the next. Following from Proposition \ref{prop: Ball of initial conditions}, it is possible to switch to the waypoint $k-1$ as soon as
\[
\begin{array}{c}
\left(\alpha\left(t\right)-\bar{\alpha}_{k-1}\right)^{2}+\dot{\alpha}^{2}\left(t\right)\leq\Delta x_{\alpha}^{2}\left(k-1\right)\\
\left(\theta\left(t\right)-\bar{\theta}_{k-1}\right)^{2}+\dot{\theta}^{2}\left(t\right)\leq\Delta x_{\theta}^{2}\left(k-1\right).
\end{array}
\]
Please note the resulting reference is
piecewise constant and the change of reference is equivalent to a
re-initialization of the continuous-time system.

\section{SIMULATIONS}\label{sec: SIMULATIONS}
\begin{figure}
\includegraphics[scale=0.64]{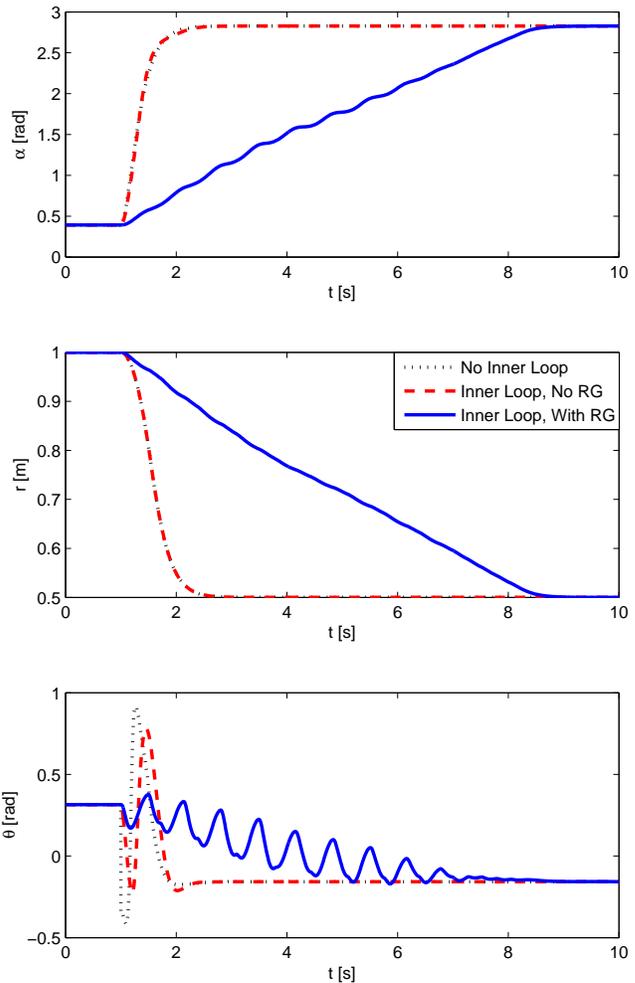}
\caption{\label{fig: Dynamics}System evolution during the numerical experiment.}
\end{figure}

\begin{figure}
\includegraphics[scale=0.64]{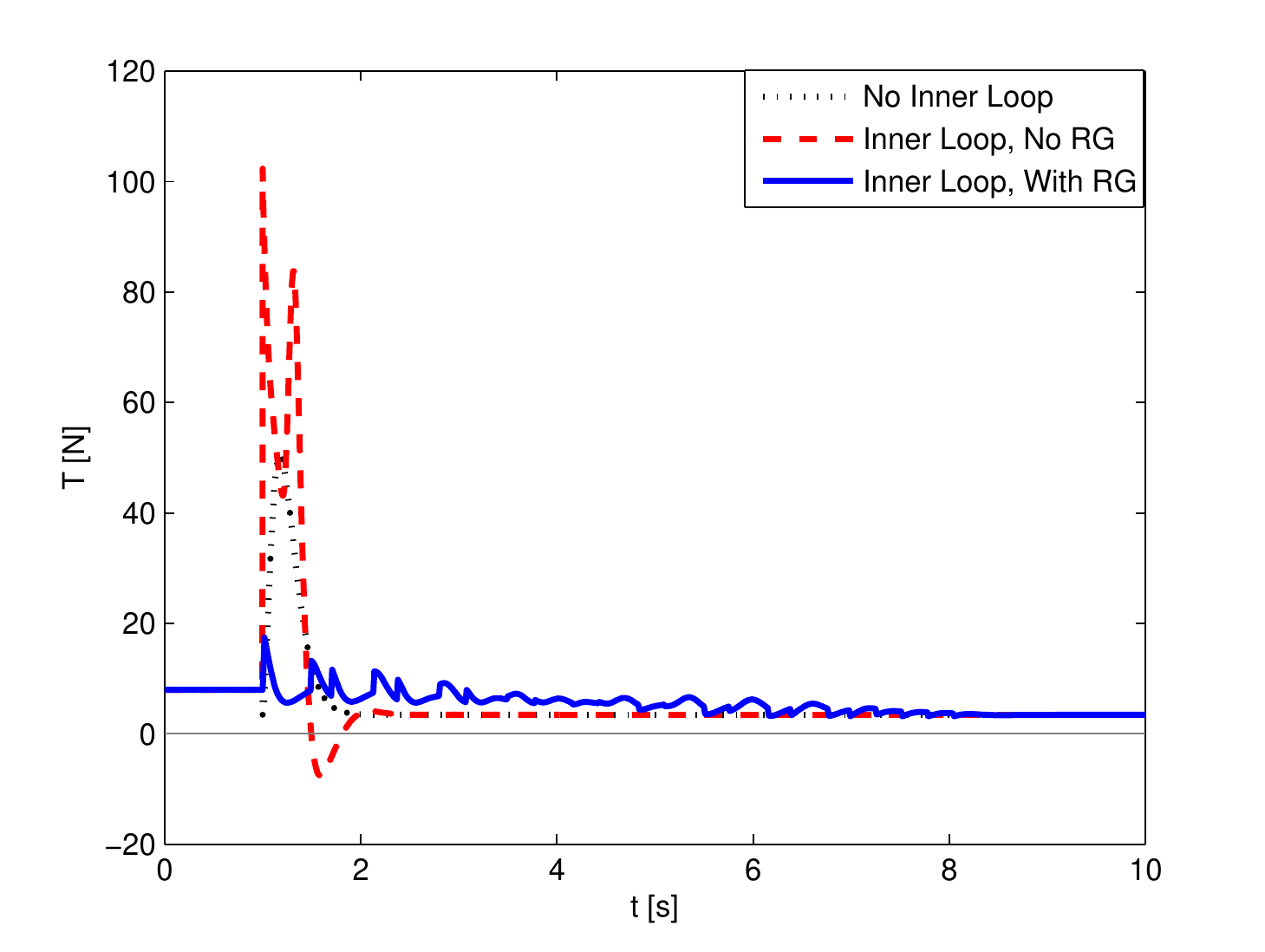}
\caption{\label{fig: Tension}Cable tension during the numerical experiment.}
\end{figure}

Consider a planar UAV of mass $m=2\,[kg]$ and moment of inertia $\mathcal{J}=0.015\,[kg\, m^{2}]$
attached to a winch of radius $\rho=0.1\,[m]$. The system is subject
to the control law (\ref{eq: u1}), (\ref{eq: u2}), (\ref{eq: u3})
and (\ref{eq: thetaC}). The outer loop gains $k_{Pr}=k_{P\alpha}=30$ have been assigned under the assumption that the inner loop is ideal. The inner loop gain $k_{P\theta}=200$ was instead chosen sufficiently high to ensure the stability of the interconnected loops. The damping factor $\zeta=0.9$ was chosen for all the derivative terms. The tethered UAV must be brought from its current
configuration $r\left(0\right)=1\,[m]$, $\alpha\left(0\right)=\frac{\pi}{8}$
and $\theta\left(0\right)=\frac{\pi}{10}$ to the desired reference
$\bar{r}=0.5\,[m]$, $\bar{\alpha}=\frac{9\pi}{10}$ and $\bar{\theta}=-\frac{\pi}{20}$.
Figure \ref{fig: Dynamics} illustrates the evolution of the elevation angle $\alpha(t)$,
the radial position $r(t)$ and the attitude angle $\theta(t)$.
Figure \ref{fig: Tension} depicts the evolution of the cable tension $T\left(t\right)$.
The simulations provide the behavior of three different control loops:
\begin{itemize}
\item \textbf{No Inner Loop:} The system response is simulated in the absence of
an attitude error (i.e. $\tilde\theta(t)=0$).
\item \textbf{Inner Loop, No RG:} The inner loop control is implemented
without the reference governor.
\item \textbf{Inner Loop, With RG:} The closed-loop system is augmented with the
Reference Governor detailed in Section \ref{sec: REFERENCE GOVERNOR}.
\end{itemize}


As illustrated in Figures \ref{fig: Dynamics}-\ref{fig: Tension}, in the absence of an attitude error the system dynamics asymptotically tend to the desired setpoint and do not violate the taut cable constraint. In the presence of the inner loop, the system has a similar dynamic response. However, the presence of an attitude error causes the violation of the taut cable constraint at time $t=1.5 [s]$. The introduction of a Reference Governor is instead able to enforce the taut cable constraint at all times, even in the presence of a non-ideal inner loop. Although the dynamic response is slower than the previous cases, it is interesting to note that the Reference Governor has the added effect of greatly reducing the maximal cable tension that is reached during the transient.

\section{CONCLUSIONS}

This paper provides a novel approach for the study of tethered UAVs
in the taut cable configuration. The cable tension is modeled as a
reaction force caused by a mechanical constraint. The system dynamics
are then obtained under the hypothesis that the taut cable condition
is verified at all times. The attainable equilibrium points are discussed
and interpreted geometrically. An inner/outer loop control strategy
is developed with the dual objective of controlling the UAV and guaranteeing
the taut cable condition. The outer loop is designed to automatically
satisfy the constraints given under the assumption of an ideal inner
loop. The inner loop error dynamics are then accounted for using a
reference governor to avoid constraint violation. Future work will
aim at the extension to the three-dimensional case as well as the
investigation of a more sophisticated reference governor strategy
to improve the system response.

\bibliographystyle{plain}
\bibliography{../../CAT-AVIATOR}

\appendix
\subsection{Proof of Proposition \ref{prop: Radial Control}}

For the sake of simplicity, the upper and lower bound of $r\left(t\right)$ will be proven only for the case $r\left(0\right)\geq\bar{r}$.\\
The initial conditions for the system are divided in four possible cases: \\
\textbf{1-} $0<\dot{r}\left(\tau_1\right)\leq\lambda_1/k_{Dr}$. Let $\tau_1$ and $\tau_{2}$ be such that $\dot{r}\left(t\right)<0$, $\forall t\in\left[\tau_1,\tau_2\right)$. In this time period, it follows that
\begin{equation}\label{eq: Lower Bound}
\ddot{r}\!=\!-\sigma_{\lambda_{1}}\left(k_{Dr}\dot{r}\!+\!\sigma_{\lambda_{2}}\left(k_{Pr}\left(r-\bar{r}\right)\right)\right)\leq-\sigma_{\lambda_{1}}\left(k_{Dr}\dot{r}\right).
\end{equation}
Thus implying that the trajectories of (\ref{eq: Lower Bound}) can be upper-bounded by the solution of $\ddot{r}=-k_{Dr}\dot{r}$. Therefore, conditions
\[
\dot{r}\left(t\right)\leq \dot{r}\left(\tau_1\right)\mathrm{e}^{-k_{Dr}t}\leq \dot{r}\left(\tau_1\right),
\]
\[
r\left(t\right)\leq r\left(\tau_1\right)+\frac{1}{k_{Dr}}\dot{r}\left(\tau_1\right)\left(1-\mathrm{e}^{-k_{Dr}t}\right)\leq r\left(\tau_1\right)+\frac{\dot{r}\left(\tau_1\right)}{k_{Dr}}
\]
and
\[
r\left(\tau_1\right)\leq r\left(t\right)
\]
hold true $\forall t\in\left[\tau_1,\tau_2\right)$.\\
Due to the presence of the term $-\sigma_{\lambda_{2}}(k_{Pr}\left(r-\bar{r}\right))$, there exists a finite time $\tau_2$ at which $\dot{r}\left(\tau_2\right)=0$. At this time instant, future trajectories can be studied by re-initializing the system in cases 2 or 3. \\
\textbf{2-} $-\lambda_1/k_{Dr}\leq\dot{r}\left(\tau_2\right)\leq0$ and $r(\tau_2)-\bar{r}>\lambda_2/k_{Pr}$. Let $\tau_2$ and $\tau_{3}$ be such that $\dot{r}\left(t\right)<0$ and $r(t)-\bar{r}>\lambda_2/k_{Pr}$, $\forall t\in\left[\tau_2,\tau_3\right]$. In this time period,
\[
\ddot{r}=-\sigma_{\lambda_{1}}\left(k_{Dr}\dot{r}+\lambda_{2}\right).
\]
As a result, the system asymptotically tends to the condition $\dot{r}=-\lambda_2/k_{Dr}$. Since $\lambda_2>\lambda_1$, conditions
\[
\left\vert\dot{r}(t)\right\vert\leq\frac{\lambda_2}{k_{Dr}}.
\]
\[
r(t)\leq r(\tau_2)
\]
\[
r(t)\geq \bar{r}+\frac{\lambda_2}{k_{Pr}}
\]
hold true $\forall t\in\left[\tau_2,\tau_3\right]$.\\
Since $\dot{r}$ asymptotically tends to a negative value, there exists a finite time $\tau_3$ at which $r(\tau_3)=\bar{r}+\lambda_2/k_{Pr}$. At time $\tau_3$, the system will always satisfy the requirements of case 3.\\
\textbf{3-} $-\lambda_2/k_{Dr}\leq\dot{r}\left(\tau_3\right)\leq0$ and $\sqrt{k_{Pr}}\tilde{r}(\tau_3)+\dot{r}(\tau_3)\geq0$. Since the saturation functions $\sigma_{\lambda1}$ and $\sigma_{\lambda2}$ are not active at time $\tau_3$, consider the dynamics of the linear system
\[
\ddot{r}=-k_{Dr}\dot{r}-k_{Pr}\left(r-\bar{r}\right)
\]
initialized in these conditions. Following from the standard linear systems theory, for $k_Dr=2\sqrt{k_{Pr}}$, the trajectory satisfies
\begin{equation}\label{eq: Trajectory}
r(t)=\bar{r}+\tilde{r}(\tau_3)e^{-\sqrt{k_{Pr}}t}+(\sqrt{k_{Pr}}\tilde{r}(\tau_3)+\dot{r}(\tau_3))te^{-\sqrt{k_{Pr}}t}
\end{equation}
for $t\in[\tau_3,\infty)$. During this whole time period, trajectory \eqref{eq: Trajectory} will never activate the saturation functions. By studying the local minima of equation
\eqref{eq: Trajectory}, it can be shown that $\sqrt{k_{Pr}}\tilde{r}(\tau_3)+\dot{r}(\tau_3)\geq0$ is a necessary and sufficient condition to ensure the absence of overshoot. As a result,
\[
|\dot{r}(t)|\leq\dot{r}(\tau_3)
\]
\[
\bar{r}\leq r(t)\leq r(\tau_3)
\]
hold true $\forall t\in\left[\tau_3,\infty\right]$.\\
The final case left to consider concerns what happens when the initial conditions will lead to an overshoot.\\
\textbf{4-} $-\lambda_1/k_{Dr}\leq\dot{r}\left(\tau_4\right)<0$ and $\sqrt{k_{Pr}}\tilde{r}(\tau_4)+\dot{r}(\tau_4)<0$. As in the previous case, the system trajectory is
\[
r(t)=\bar{r}+\tilde{r}(\tau_4)e^{-\sqrt{k_{Pr}}t}+(\sqrt{k_{Pr}}\tilde{r}(\tau_4)+\dot{r}(\tau_4))te^{-\sqrt{k_{Pr}}t}
\]
for $t\in[\tau_4,\infty)$. This time, however, the system trajectory presents a local minima at time
\[
\tau^*=\tau_4+\frac{\dot{r}(\tau_4)}{k_{Pr}\tilde{r}(\tau_4)+\sqrt{k_{Pr}}\dot{r}(\tau_4)},
\]
thus leading to the maximum overshoot
\[
r^*=\bar{r}+\left(\tilde{r}(\tau_4)+\frac{\dot{r}(\tau_4)}{\sqrt{k_{Pr}}}\right)e^{-\sqrt{k_{Pr}}\tau^*}.
\]
The proof is concluded by combining the properties of all four cases and doing an analogous study for $r\left(0\right)<\bar{r}$.

\subsection{Proof of Proposition \ref{prop: IOS Outer Loop}}
Define $x_{\alpha}=\left[\alpha-\bar{\alpha},\dot{\alpha}\right]^{T}$. The
state space expression of system (\ref{eq: TanDyn_Real}) is
\[
\left\{ \begin{array}{l}
\dot{x}_{\alpha1}=x_{\alpha2}\\
\dot{x}_{\alpha2}=-f_\alpha\left(x_{\alpha1},x_{\alpha2}\right)+\Gamma\left(r,\ddot{r},\tilde\alpha,\tilde{\theta}\right)
\end{array}\right.
\]
where
\[
f_\alpha\!\left(x_{\alpha1},x_{\alpha2}\right)\!=\!\left(k_{P\alpha}x_{\alpha1}+k_{D\alpha}x_{\alpha2}\right)\!\cos\tilde{\theta}-2\frac{\dot{r}}{r}\!\left(1-\cos\tilde{\theta}\right)\!x_{\alpha2}
\]
represents the state-dependent dynamics whereas
\[
\Gamma\left(r,\ddot{r},\tilde\alpha,\tilde{\theta}\right)=\frac{g}{r}\left(\cos\tilde{\theta}-1\right)\cos\left(\tilde{\alpha}+\bar{\alpha}\right)-\frac{u_{T}}{mr}\sin\tilde{\theta}.
\]
can be seen as an exogenous bounded input since $\left\vert\cos\left(\tilde{\alpha}+\bar{\alpha}\right)\right\vert\leq1$. To prove ISS, the first step will be the identification of a strict Lyapunov function in the condition $\Gamma=0$. To this end, define $k_{P\alpha}=\omega_{\alpha}^{2}$ $k_{D\alpha}=2\zeta \omega_{\alpha}$ and consider the Candidate Lyapunov function
\[
V=\frac{1}{2}x_{\alpha}^{T}\left[\begin{array}{cc}
h_{p}+qh_{d} & q\\
q & 1
\end{array}\right]x_{\alpha}
\]
where $h_{p}=\omega_{\alpha}^{2}\tilde{\theta}_{\max}$,
$h_{d}=2\zeta \omega_{\alpha}\tilde{\theta}_{\max}$
and $q\in\left(0,h_{d}\right)$. The time derivative is
\[
\begin{array}{l}
\dot{V}=\!-q\omega_{\alpha}^{2}\cos\tilde{\theta}x_{\alpha1}^{2}-\!2\left(\zeta \omega_{\alpha}\cos\tilde{\theta}+\frac{\dot{r}}{r}\left(1-\cos\tilde{\theta}\right)\!-q\right)\!x_{\alpha2}^{2}\\
+\!\left(\!\left(\omega_{\alpha}^{2}+2q\zeta \omega_{\alpha}\right)\!\left(\!\tilde{\theta}_{\max}\!\!-\cos\tilde{\theta}\right)\!+\!2q\frac{\dot{r}}{r}\left(\!1\!-\!\cos\tilde{\theta}\right)\!\right)\!x_{\alpha1}x_{\alpha2}
\end{array}
\]
which is upper-bounded by
\[
\dot{V}\leq x_{\alpha}^TQx_\alpha
\]
where
\[
Q=\left[\begin{array}{cc}
Q_{11} & Q_{12}\\
Q_{12} & Q_{22}
\end{array}\right]
\]
and
\[
\begin{array}{l}
Q_{11}=q\omega_{\alpha}^{2}\tilde{\theta}_{\max}\\
Q_{12}=\frac{1}{2}\left(\omega_{\alpha}^{2}+2q\left(\zeta \omega_{\alpha}-\left\Vert \frac{\dot{r}}{r}\right\Vert _{\infty}\right)\right)\left(1-\tilde{\theta}_{\max}\right)\\
Q_{22}=2\zeta \omega_{\alpha}\tilde{\theta}_{\max}-2\left\Vert \frac{\dot{r}}{r}\right\Vert _{\infty}\left(1-\tilde{\theta}_{\max}\right)-q
\end{array}
\]
As a result, it follows that $Q<0\Rightarrow\dot{V}<0$. To obtain a negative-definite $Q$ it is necessary to impose $q>0$ and ensure
\begin{equation}\label{eq: QmatrixCondition}
\begin{array}{l}
q\omega_{\alpha}^{2}\tilde{\theta}_{\max}\!\!\left(2\zeta \omega_{\alpha}\tilde{\theta}_{\max}\!\!\!-\!2\left\Vert \frac{\dot{r}}{r}\right\Vert _{\infty}\!\!\left(1\!-\!\tilde{\theta}_{\max}\right)\!-\!q\right)\\
>\frac{1}{4}\left(\omega_{\alpha}^{2}+2q\left(\zeta \omega_{\alpha}-\left\Vert \frac{\dot{r}}{r}\right\Vert _{\infty}\right)\right)^{2}\left(1-\tilde{\theta}_{\max}\right)^{2}.
\end{array}
\end{equation}
By parameterizing
\begin{equation}\label{eq: Restriction R}
\left\Vert \frac{\dot{r}}{r}\right\Vert _{\infty}=\nu\frac{\zeta \omega_{\alpha}\tilde{\theta}_{\max}}{\left(1-\tilde{\theta}_{\max}\right)}
\end{equation}
with $\nu\in\left(0,1\right)$, equality (\ref{eq: QmatrixCondition}) becomes
\[
\begin{array}{l}
q\omega_{\alpha}^{2}\tilde{\theta}_{\max}\left(2\zeta \omega_{\alpha}\tilde{\theta}_{\max}\left(1-\nu\right)-q\right)\\
>\frac{1}{4}\left(\omega_{\alpha}^{2}+2q\zeta \omega_{\alpha}\left(1-\nu\right)\right)^{2}\left(1-\tilde{\theta}_{\max}\right)^{2}.
\end{array}
\]
At this point, the parameter $q\in(0,h_d)$ can be chosen so as to maximise the term to the left of the inequality. Thus, by choosing
\[
q=\zeta \omega_{\alpha}\tilde{\theta}_{\max}\left(1-\nu\right)
\]
equality (\ref{eq: QmatrixCondition}) becomes
\[
a\omega_{\alpha}^{4}\tilde{\theta}_{\max}^{3}>b\omega_{\alpha}^{4}\left(1-\tilde{\theta}_{\max}\right)^{2}
\]
with
\[
\begin{array}{l}
a=\left(1-\nu\right)^{2}\zeta^{2}\\
b=\frac{1}{4}\left(1+2\left(1-\nu\right)^{2}\zeta^{2}\right)^{2}.
\end{array}
\]
At this point, the value of $\left\Vert \tilde{\theta}\right\Vert _{\infty}$ which guarantees $Q<0$ can then be obtained by solving
\begin{equation}\label{eq: Restriction th}
a\tilde{\theta}_{\max}^{3}-b\left(1-\tilde{\theta}_{\max}\right)^{2}>0.
\end{equation}
Having obtained a negative definite $\dot{V}$, consider what happens if $\Gamma\neq0$. Following the same reasoning as before, the derivative of $V(x_\alpha)$ is lower bounded by
\[
\dot{V}\leq-x_{\alpha}^{T}Qx_{\alpha}+x_{\alpha}^TR\Gamma
\]
where
\[
R = \left[\begin{array}{c} q\\ 1\end{array}\right].
\]
To prove ISS, it is sufficient to note that
\[
x_{\alpha}^TR\Gamma\leq\left\Vert x_\alpha \right\Vert  \left\Vert R\Gamma \right\Vert
\]
and
\[
x_{\alpha}^{T}Qx_{\alpha}\geq\underline{\lambda}_Q\left\Vert x_\alpha \right\Vert^2
\]
where $\underline{\lambda}_Q$ is the lowest eigenvalue of the positive definite matrix $Q$. As a result,
\[
\left\Vert x_\alpha \right\Vert\geq\left\Vert R\Gamma \right\Vert
\]
implies $\dot{V}\leq0$, thus proving ISS whenever $\tilde\theta$ and $\dot{r}/r$ satisfy inequalities (\ref{eq: Restriction R})-(\ref{eq: Restriction th}).

\subsection{Proof of Proposition \ref{prop: IOS Inner Loop}}
Define $x_{\theta}=\left[\tilde{\theta},\dot{\theta}\right]^{T}$ and
$q=\sqrt{k_{P\theta}}$. The closed loop dynamics of the inner loop are
\begin{equation}
\begin{array}{l}
\dot{x}_{\theta}=\left[\begin{array}{cc}
0 & 1\\
-q^{2} & -2\zeta q
\end{array}\right]x_{\theta}+\left[\begin{array}{c}
1\\
0
\end{array}\right]\dot{\theta}_{C}\\
\;\;\tilde\theta=\left[\begin{array}{cc}
1 & 0\end{array}\right]x_{\theta}
\end{array}.\label{eq: Beta State Space}
\end{equation}
Since the state matrix is Hurwitz, the system is ISS \cite{ISS_BasicConcepts}.
The asymptotic gain $\gamma_{\mathrm{In}}$ between the input $\dot\theta_C$ and
the output $\tilde{\theta}$ is the $\ell_{1}$ norm
\begin{equation}\label{eq: l1 norm_1}
\gamma_{\mathrm{In}}=\int_{0}^{\infty}\left\vert C e^{As}B\right\vert \mathrm{d}s
\end{equation}
Given $\zeta\in\left(0,1\right)$, the two eigenvalues of the state matrix $A$
are complex-conjugate. The matrix exponential can therefore be re-written as
\[
e^{At}=\left(2\Sigma\cos\omega t-2\Omega\sin\omega t\right)e^{\sigma t},
\]
where
\[
\begin{array}{ll}
\sigma=\mathrm{Re}\left(\lambda_{1}\right) & \omega=\mathrm{Im}\left(\lambda_{1}\right)\\
\Sigma=\mathrm{Re}\left(v_1w_1^T\right) & \Omega=\mathrm{Im}\left(v_1w_1^T\right)
\end{array}
\]
and $\lambda_1$ is the first eigenvalue of the state matrix whereas $v_1$ and $w_1$ are
the corresponding left and right eigenvectors. As a result,
\[
e^{At}=\left(\!\left[\!\begin{array}{cc}
1 & 0\\
0 & 1
\end{array}\!\right]\!\cos\omega t+\frac{1}{\sqrt{1-\zeta^{2}}}\!\left[\!\begin{array}{cc}
-\zeta & -\frac{1}{q}\\
q & \zeta
\end{array}\!\right]\!\sin\omega t\!\right)\! e^{-q\zeta t}
\]
and
\begin{equation}\label{eq: l1 norm_2}
Ce^{As}B=\left(\cos\omega t-\frac{\zeta}{\sqrt{1-\zeta^{2}}}\sin\omega t\right)e^{-q\zeta t}.
\end{equation}
By combining (\ref{eq: l1 norm_1})-(\ref{eq: l1 norm_2}), it follows that
\[
\gamma_{\mathrm{In}}=\int_{0}^{\infty}\left\vert \cos(\omega s)-\frac{\zeta}{\sqrt{1-\zeta^{2}}}\sin(\omega s)\right\vert e^{-q\zeta s}\mathrm{d}s.
\]
For any $\zeta\in(0,1)$, the following upper bound applies
\[
\gamma_{\mathrm{In}}\leq\frac{1}{\zeta q}.
\]

\end{document}